\documentclass[longauth]{aa}

\usepackage{txfonts}
\usepackage{graphicx,times}
\usepackage{natbib}
\usepackage{amssymb,amsmath}
\bibpunct{(}{)}{;}{a}{}{,}
\usepackage{epstopdf}

\defcitealias{gao_2004_a}{2004a}
\defcitealias{gao_2004_b}{2004b}
\defcitealias{melioli2013}{Melioli, de Gouveia Dal Pino, \& Geraissate 2013}
\defcitealias{Braine_2017}{2017}
\defcitealias{Braine_2023}{2023}
\defcitealias{Li_2020}{2020}
\defcitealias{Lifei_2021}{2021}

\usepackage{array}
\usepackage{float}
\usepackage{multirow}
\usepackage{rotating}
\usepackage{csvsimple}
\usepackage[pagebackref=false]{hyperref}
\usepackage{threeparttable}
\usepackage[makeroom]{cancel}

\usepackage{orcidlink} 
\usepackage[normalem]{ulem}

\newcommand{\Jt}[2]{$J =$ {#1}$-${#2}}

\usepackage{titlesec}

\makeatletter
\let\do@linenumbers\relax
\let\linenumbers\relax
\let\endlinenumbers\relax
\makeatother

\begin{document}

	\title{The MALATANG survey: Dense gas distribution on sub-kiloparsec scales across the disk of M82}

\author{
	Jian-Fa Wang\orcidlink{0000-0001-6693-0743}\inst{1,2},
	Yu Gao\orcidlink{0000-0003-0007-2197}\inst{3}\thanks{Deceased}, 
	Qing-Hua Tan\orcidlink{0000-0003-3032-0948}\inst{1}, 
	Xue-Jian Jiang\orcidlink{0000-0002-8899-4673}\inst{4}, 
	Li Ji \inst{1,5}, 
	Zhi-Yu Zhang\orcidlink{0000-0002-7299-2876}\inst{6,7},  
	Jun-Zhi Wang\inst{8}, 
	Jun-Feng Wang\inst{3}, 
	Thomas R. Greve\orcidlink{0000-0002-2554-1837}\inst{9,10}, 
	Yan Jiang\inst{11}, 
	Ashley Bemis\inst{12},
	Elias Brinks\orcidlink{0000-0002-7758-9699}\inst{13}, 
	Aeree Chung\inst{14}, 
	Malcolm J. Currie\orcidlink{0000-0003-0144--3362}\inst{15,16}, 
	Richard de Grijs\orcidlink{0000-0002-7203-5996}\inst{17,18,19},  
	Taotao Fang\orcidlink{0000-0002-2853-3808}\inst{3},
	Luis C. Ho\orcidlink{0000-0001-6947-5846}\inst{20,21}, 
	Bumhyun Lee\inst{14}, 
	Satoki Matsushita\orcidlink{0000-0002-2127-7880}\inst{22}, 
	Micha{\l} Micha{\l}owski\inst{23}, 
	Soojong Pak\orcidlink{0000-0002-2548-238X}\inst{24}, 
	Panomporn Poojon\inst{14,25}, 
	Mark G. Rawlings\orcidlink{0000-0002-6529-202X}\inst{26}, 
	Amelie Saintonge\orcidlink{0000-0003-4357-3450}\inst{27}, 
	Yi-Chen Sun\orcidlink{0000-0003-0477-1606}\inst{6,7}, 
	Jing Zhou\inst{6,7}
}

\institute{
	Purple Mountain Observatory, Chinese Academy of Sciences, No.10 Yuanhua Road,  Qixia District, Nanjing 210023, China; {\it jfwang@pmo.ac.cn; qhtan@pmo.ac.cn}
	\and
	University of Science and Technology of China, No.96, JinZhai Road Baohe District, Hefei, Anhui 230026, China
	\and
	Department of Astronomy, Xiamen University, 422 Siming South Road, Xiamen, 361005, China
	\and
	Research Center for Astronomical Computing, Zhejiang Laboratory, Hangzhou 311100, China
	\and
	Key Laboratory of Dark Matter and Space Astronomy, CAS, Nanjing 210023, China
	\and
	School of Astronomy and Space Science, Nanjing University, Nanjing 210093, China
	\and
	Key Laboratory of Modern Astronomy and Astrophysics (Nanjing University), Ministry of Education, Nanjing 210093, China
	\and
	Guangxi Key Laboratory for Relativistic Astrophysics, School of Physical Science and Technology, Guangxi University, Nanning 530004,  China
	\and
	Cosmic Dawn Center (DAWN), Denmark
	\and
	DTU-Space, Technical University of Denmark, Elektrovej 327, 2800 Kgs. Lyngby, Denmark
	\and
	School of Physics and Astronomy, China West Normal University, No. 1 Shida Road, Nanchong 637002, China
	\and
	Department of Physics and Astronomy, McMaster University, Hamilton, ON L8S 4M1, Canada
	\and
	Centre for Astrophysics Research, University of Hertfordshire, College Lane, Hatfield AL10 9AB, UK
	\and
	Department of Astronomy, Yonsei University, 50 Yonsei-ro, Seodaemun-gu, Seoul 03722, Republic of Korea
	\and
	RAL Space, Rutherford Appleton Laboratory, Harwell Campus, Didcot OX11 0QX, UK
	\and
	East Asian Observatory, 660 N. A‘ohōkū Place, Hilo, HI 96720, USA
	\and
	School of Mathematical and Physical Sciences, Macquarie University, Balaclava Road, Sydney NSW 2109, Australia
	\and
	Astrophysics and Space Technologies Research Centre, Macquarie University, Balaclava Road, Sydney, NSW 2109, Australia
	\and
	International Space Science Institute--Beijing, 1 Nanertiao, Zhongguancun, Hai Dian District, Beijing 100190, China
	\and
	Kavli Institute for Astronomy and Astrophysics, Peking University, Beijing 100871, China
	\and
	Department of Astronomy, School of Physics, Peking University, Beijing 100871, China
	\and
	Institute of Astronomy and Astrophysics, Academia Sinica,  No.1, Sec. 4, Roosevelt Rd, Taipei 10617, Taiwan
	\and
	Astronomical Observatory Institute, Faculty of Physics and Astronomy, Adam Mickiewicz University, ul.~S{\l}oneczna 36, 60-286 Pozna{\'n}, Poland
	\and
	School of Space Research, Kyung Hee University, 1732 Deogyeong-daero, Yongin 17104, Korea
	\and
	National Astronomical Research Institute of Thailand, 260 Moo 4, Donkaew, Mae Rim, Chiang Mai, 50180, Thailand
	\and
	Gemini Observatory/NSF NOIRLab, 670 N. A'ohoku Place, Hilo, Hawai'i, 96720, USA
	\and
	Department of Physics \& Astronomy, University College London, Gower Street, London, WC1E 6BT, UK
}

	\abstract{We present observations of HCN \Jt{4}{3} and HCO$^+$ \Jt{4}{3} lines obtained with the James Clerk Maxwell Telescope as part of the  MALATANG survey, combined with archival HCN \Jt{1}{0} and HCO$^+$ \Jt{1}{0}  data from the Green Bank Telescope, to study the spatial distribution and excitation conditions of dense molecular gas in the disk of M82. We detect HCN \Jt{4}{3} and HCO$^+$ \Jt{4}{3} emission within the central region ($\lesssim$ 500 pc) of the galaxy, while the \Jt{1}{0} emission lines exhibit a more extended spatial distribution ($\gtrsim$ 700 pc).
		The dense gas shows a clear double-lobed structure in both spatial distribution and kinematics, with the HCN and HCO$^+$ \Jt{4}{3} lines in the southwest lobe blueshifted by $\sim$ 40 km s$^{-1}$ relative to the \Jt{1}{0} lines.
		The HCN \Jt{4}{3}/1$-$0 and HCO$^+$ \Jt{4}{3}/1$-$0 line-luminosity ratios range from 0.09 to 0.53 and from 0.14 to 0.87, respectively, with mean values of 0.18 $\pm$ 0.04 and 0.36 $\pm$ 0.06. 
		The HCN ratio is lower than the typical average observed in nearby star-forming galaxies, whereas the HCO$^+$ ratio is comparatively higher, suggesting that the high-$J$ HCN emission in M82 is significantly sub-thermally excited.
		Spatially, the peak values of the \Jt{4}{3}/1$-$0 ratios are found in the northwest region of M82, coinciding with the galaxy-scale outflow. Elevated HCN/HCO$^+$ ratios are also detected in roughly the same area, potentially tracing local excitation enhancements driven by the outflow.
		The HCN/HCO$^+$  \Jt{4}{3} ratio across all detected regions ranges from 0.19 to 1.07 with a mean value of 0.41 $\pm$ 0.11, which is significantly lower than the average \Jt{1}{0} ratio of 0.76 $\pm$ 0.08.
		Both ratios are significantly lower than the average values observed in nearby star-forming galaxies, which could be related to the relatively low gas density and the presence of an extended photo-dissociation region in M82.
	}

	\titlerunning{Dense gas distribution on sub-kiloparsec scales across the disk of M82}
	\authorrunning{Jian-Fa Wang et al.}

\maketitle

\newpage
	\section{Introduction}

Molecular gas serves as a reservoir of raw material for star formation. Numerous studies have demonstrated that star formation in galaxies is more directly linked to dense gas rather than to diffuse atomic gas or low-density molecular gas components (e.g., Gao \& Solomon \citetalias{gao_2004_a,gao_2004_b}; \citealt{Gao_2007}; \citealt{baan2008}; \citealt{Bigiel_2008}; \citealt{liu_gao2010}; \citealt{Wang_2011}; \citealt{garcia_2012}; \citealt{Kennicutt_2012}; \citealt{zhang2014}; \citealt{liu2016}; \citealt{onus2018}; Li et al. \citetalias{Li_2020}; \citetalias{Lifei_2021}).

M82 is a prototypical starburst galaxy, characterized by a galactic-scale outflow (initially observed by \citealt{Lynds_1963}). 
With properties listed in Table~\ref{tab:properties}, it provides an excellent environment for investigating the connection between star formation and gas in a galaxy, owing to its proximity ($D \approx 3.5$ Mpc; \citealt{Dalcanton_2009}), elevated infrared luminosity ($L_\mathrm{IR} \approx 5.6 \times 10^{10} L_{\odot}$; \citealt{Sanders_2003}), and substantial star formation rate ($\sim$ 9 M$_{\odot}$ yr$^{-1}$ over the last 100 Myr; e.g., \citealt{Calzetti_2013,Strickland_2004}).
Far-infrared continuum observations have revealed a characteristic dust temperature of $\sim$30 K in the central starburst region of M82 (e.g., \citealt{Pattle_2023,Roussel_2010}).
The molecular gas in M82 has been studied through a variety of CO-line observations across the millimeter to far-infrared (e.g., \citealt{Loiseau_1990,Wei_2001,walter_2002,Salak_2013,Leroy_2015,Chisholm_2016}). The molecular gas disk exhibits a prominent double-lobed structure, with bright lobes located northeast (NE) and southwest (SW) of the galaxy's center. This structure is clearly seen in low-$J$~$^{12}$CO (hereafter simply referred to as CO when not otherwise specified) and $^{13}$CO transitions, and is preserved in higher transitions up to at least \Jt{7}{6} (e.g., \citealt{loenen_2010}). While the lobes are roughly symmetric in position, the intensity distribution is asymmetric (e.g., \citealt{Kikumoto_1998}). Furthermore, the separation between the two lobes varies with $J$ number (e.g., \citealt{Mao_2000,Ward_2003}). These trends indicate the presence of two molecular gas components: low-excitation gas that is more extended and dominates low-$J$ transitions, and high-excitation gas that is more compact and dominates mid- to high-$J$ transitions (e.g., \citealt{loenen_2010}).
$^{13}$CO shows a similar overall distribution as CO but is relatively weaker and more centrally concentrated. The CO/$^{13}$CO intensity ratio is generally high in M82, possibly due to enhanced UV radiation in the starburst region that photodissociates $^{13}$CO more efficiently (e.g., \citealt{Loiseau_1988,Neininger_1998}).
It is also possible that elevated local temperatures cause the $^{13}$CO(1--0) to become optically thin by shifting the population distribution toward higher levels (e.g., \citealt{Matsushita_2010}). 
These results suggest that the molecular gas in M82 is multi-phase and structurally complex.

Low-$J$ CO transitions have low critical densities ($\sim$300--1000 cm$^{-3}$) and are often used to trace the total molecular gas as a proxy for H$_2$, which lacks an electric dipole moment (e.g., \citealt{Bolatto_2013}). Higher-$J$ CO transitions ($J$ $\geq$ 4--3) have higher critical densities, but are more sensitive to warm molecular gas, while dense-gas tracers such as HCN and HCO$^+$ are used to trace cold and dense molecular gas (e.g., \citealt{Shirley_2015}). Combining different tracers is essential for understanding the distribution and excitation of molecular gas in galaxies. Existing observations of HCN and HCO$^+$ in M82 have primarily focused on the \Jt{1}{0} transitions (e.g., \citealt{Salas_2014,Kepley_2014,Ginard_2015,Li_2022}). While detections of high-$J$ HCN and HCO$^+$ lines have been reported in the galaxy's center and its surrounding regions (e.g., \citealt{Isreal_2023,Seaquist_2006}), 
spatially resolved maps and systematic comparisons across multiple transitions remain scarce.

A major challenge in studying the distribution and excitation state of dense gas in extragalactic systems  lies in the intrinsic faintness of its emission. In most regions of galaxies,  HCN and $\rm HCO^{+}$ lines are significantly weaker than those of CO, by a factor of typically by 10 to 30 (e.g., \citealt{gao_2004_b}). 
Benefiting from the 16-receptor array receiver Heterodyne Array Receiver Program (HARP, \citealt{Buckle_2009}), the James Clerk Maxwell Telescope (JCMT) is capable of efficiently producing  large-scale maps, enabling  spatially resolved observations of dense molecular gas in nearby galaxies.
The JCMT-MALATANG program aims to map HCN(4--3) and $\rm HCO^{+}$(4--3) emission in 23 nearby galaxies (\citealt{tan2018,jiang2020}). In this study, we combine MALATANG  observations of M82,  with HCN(1--0) and $\rm HCO^{+}$(1--0) data from \cite{Salas_2014}, along with additional archival  datasets, to  investigate the spatial distribution and excitation state of dense gas across  the disk of M82.

In Sect. \ref{sect:DATA} we  briefly describe the observations and data reduction. Section \ref{sect:results} presents the spectra, stacking results, velocity distributions,  and line ratios. 
In Sect. \ref{sect:discussion}, we analyze the trends in the line ratios, explore their possible physical origins, and compare our findings with previous studies. Our main results are summarized in Section \ref{sect:conc}.

	\section{Observations and  data reduction}
	\label{sect:DATA}

	\subsection{JCMT HCN(4--3) and HCO$+$(4--3)  observations}
	
	\begin{figure}
		\centering
		\includegraphics[width=0.48\textwidth]{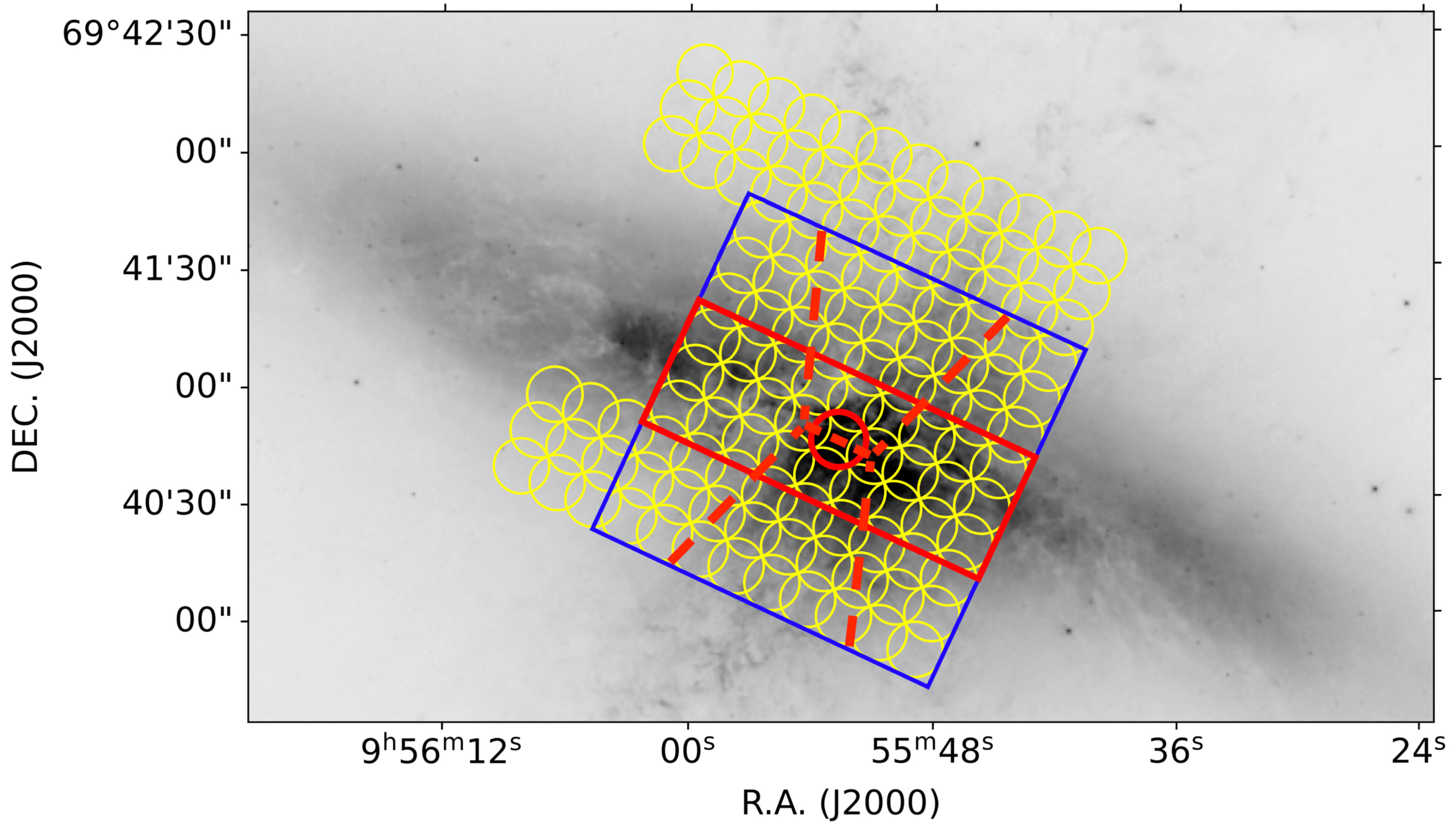}
		\caption{JCMT MALATANG observed positions for M82 overlaid on HST H$\alpha$ emission (NASA, ESA, The Hubble Heritage Team, \citealp{Mutchler_2007}). The yellow circles indicate to the spatial areas covered by the JCMT HARP receivers using jiggle mode, and the galaxy's center is marked by a red circle. The 9 $\times$ 9 JCMT MALATANG observation of  HCN(4--3) and $ \rm HCO^{+}$(4--3) is represented by the blue box. In this study, we define the region enclosed by the red rectangle as the area dominated by the disk of M82, which is slightly larger than the CO molecular disk described in \cite{walter_2002} and \cite{Salas_2014}. 
		Following the definition of the outflow regions in \cite{Leroy_2015} based on CO, H$\alpha$, FUV, and X-ray emission, we represent the regions associated with outflow as a conical structure with a base of 300 pc and an opening angle of 20$^\circ$, as indicated by the red dashed lines.
		}
		\label{fig: position}
	\end{figure}

	The first phase of the JCMT Large Program MALATANG (project code: M16AL007) conducted a total of approximately 390 hours of observations from December 2015 to July 2017, mapping HCN(4--3) and $ \rm HCO^{+}$(4--3) line emission in 23 nearby infrared-bright star-forming galaxies (\citealt{tan2018}; \citealt{jiang2020}).  
	For M82, observations of HCN(4--3) and $\rm HCO^{+}$(4--3) were conducted across the central $2^\prime \times 2^\prime$ region, which is centered at R.A. (J2000.0) = 09$^{\textrm h}$55$^{\textrm m}$52{\fs}4 and Dec (J2000.0) = +$69{\textrm \degr}40{\textrm \arcmin}46{\farcs}9$. The total integration times were 150 minutes for HCN(4--3) and 100 minutes for $\rm HCO^{+}$(4--3), including the time spent integrating on both the source and the reference position.
	The HARP array was used to carry out $3~\times~3$ jiggle-mode observations with grid spacing of $10\arcsec$.  Figure \ref{fig: position} shows the mapped positions for M82, while the incompleteness of observation coverage was due to two adjacent receptors (H13 and H14) at the edge of the array not working. 
	The receiver backend is the Auto-Correlation Spectral Imaging System (ACSIS) spectrometer with a bandwidth of 1 GHz and a resolution of 0.488 MHz, which correspond to 840 km s$^{-1}$ and 0.41 km s$^{-1}$ at 354 GHz, respectively. The FWHM beam width of each receiver at 350 GHz is about $14\arcsec$, corresponding to a physical scale of 245 pc at M82's distance of 3.5 Mpc. All the MALATANG observation parameters of M82 are shown in Table \ref{tab:obser para}.

	\begin{table}[h]
	\caption{Basic properties of M82}
	\label{tab:properties}
	\centering
	\begin{threeparttable}
		\begin{tabular}{lcl}
			\hline
			Parameters  &  Value  & Ref. \\
			\hline 
			R.A. (J2000) & 09$^{\rm h}${}55$^{\rm m}${}52{\rm \fs}4 & (1) \\ 
			Dec. (J2000) & $+$69{\textrm \degr}40{\textrm \arcmin}46{\farcs}7 & (1) \\ 
			Distance & 3.5 Mpc & (1) \\ 
			Diameter ($D_{25}$) & 11.2\arcmin $\times$ 4.3\arcmin & (2) \\
			Velocity (LSR) & 225 km s$^{-1}$ & (2) \\  
			Morphology Type & I0, edge-on & (3)  \\
			Inclination & 80$\degr$ & (3) \\
			Position angle & 65$\degr$ & (3) \\

			\hline
		\end{tabular}
		\begin{tablenotes}[flushleft]
			\item \normalfont References: 
			
			(1) \cite{Karachentsev_2004};
			
			(2) \cite{Neininger_1998};
			
			(3) \cite{deVaucouleurs1991}.

		\end{tablenotes}
	\end{threeparttable}
\end{table}

	\begin{table}
		\caption{Summary of MALATANG  observational parameters  for M82}
		
		\label{tab:obser para}
		\centering
		\begin{threeparttable}
			\begin{tabular}{lcl}
				\hline
				Parameters  &  HCN(4--3) & $\rm HCO^{+}$(4--3)  \\
				\hline
				Observation date & 2015 Dec 10, 2015 Dec 12 & 2015 Dec 13 \\ 
				Frequency    & 354.265 GHz  	    & 356.494 GHz   \\ 
				{ROT\_PA}  \tnote{+} &  65 deg & 65 deg \\  
				$\overline{T_{\rm sys}}$ &  270 K 		& 338 K	\\  
				$\overline {\tau}$ (225 GHz)   &  0.031  	& 0.051 	\\ 
				$t_{\rm int}$ \tnote{*} &  150 min     & 100 min 	\\
				Mapping region &  $2\arcmin~\times~2\arcmin$   &  $2\arcmin~\times~2\arcmin $	\\
				FWHM &  $14\arcsec$    &  $14\arcsec$ 	\\
				\hline
			\end{tabular}
			\begin{tablenotes}[flushleft]
				\item \normalfont +  Position angle of the galaxy's major axis,  adopted for mapping the HCN and HCO$^+$ emission.
				\item \normalfont *  Total integration time including ON + OFF.
			\end{tablenotes}
		\end{threeparttable}	
	\end{table}

	\subsection{GBT HCN(1--0), HCO$^+$(1--0) and  ancillary data}

	The HCN(1--0) and $ \rm HCO^{+}$(1--0) data were obtained from \cite{Kepley_2014} and \cite{Salas_2014}, observed with the 4-mm (W band) receiver and spectrometer of the Green Bank Telescope (GBT). Observations were conducted under excellent weather conditions for approximately 15 hours. The beam size of the calibrated cubes is 9\farcs2. In order to compare these with the HCN(4--3)  and $ \rm HCO^{+}$(4--3) data, we convolved the GBT data to a $14\arcsec$ resolution, following the method of \cite{Aniano_2011}. 
		
	We obtained the M82 ${}^{12}\mathrm{CO}$(1--0) data from the Nobeyama COMING project\footnote{https://astro3.sci.hokudai.ac.jp/~radio/coming/} (\citealp{Sorai_2019}). At 115 GHz, the beamsize of the Nobeyama 45-m radio telescope is approximately $14\arcsec$, which is consistent with the MALATANG data. The CO(3--2) data were obtained from the JCMT Nearby Galaxies Survey (\citealp{Wilson_2012}).

	\subsection{Data reduction}

	This work used the GILDAS/CLASS\footnote{http://www.iram.fr/IRAMFR/GILDAS/} software package for spectral reduction throughout the entire process.  
	These JCMT NDF ($N$-Dimensional Data Format) files were converted into a format compatible with the GILDAS/CLASS software package, retaining all the raw spectra.
	We filtered each individual raw spectrum instead of the averaged results at a given position, allowing for more spectral data to be retained. First, we manually inspected and removed spectra with obvious issues. Then, we developed an automated script to mask the 10\% of spectra with the highest noise levels. 
	Next, we averaged the spectra at each position and combined data from different scans using noise-weighted averaging.
	After these steps, we performed low-order-polynomial baseline subtraction and used the velocity range over which line emission was detected in CO(3--2) as a reference for where we expect emission to be found for the other observed lines, and to calculate upper limits. 
	Finally, we smoothed the spectra to a velocity resolution of approximately 13 km\ s$^{-1}$. Figure \ref{fig: A1} presents a comparison of selected spectra from this work and \cite{tan2018}.

	\subsection{Measurements of flux density and line luminosities}
	
	We adopted the same method as proposed by \cite{Gao_1996} to identify the emission lines, based on the criterion that the velocity-integrated line intensity $ \geqslant 3\sigma$, where the uncertainty $\sigma$ is obtained from the following formula:
	\begin{equation}
		\sigma_I = T_{\text{rms}} \sqrt{\Delta v_{\text{line}} \Delta v_{\text{res}}} \sqrt{1+ \frac{\Delta v_{\text{line}}}{\Delta v_{\text{base}}}}.
	\end{equation}

	Here, $T_{\rm rms}$ denotes the root-mean-square (RMS) main-beam temperature of the line data, considering a spectral velocity resolution of $\Delta v_{\rm res}$, $\Delta v_{\rm line}$ represents the velocity range of the emission line, while $\Delta v_{\rm base}$ is the velocity range employed to fit the baseline.
	For the JCMT HCN(4--3), $ \rm HCO^{+}$(4--3) and CO(3--2) data, the flux density is calculated using $S/T_{\rm mb}=15.6/\eta_{\rm mb}=24.4\ {\rm Jy\ K^{-1}}$  (\citealp{tan2018} for more details). For the GBT data, we adopt the main-beam efficiency \(\eta_{\text{mb}} = 0.26\) (\citealp{Kepley_2014}) and $S/T_{\rm mb}\footnote{https://www.gb.nrao.edu/GBT/Performance/PlaningObservations.htm} = 1\ {\rm Jy\ K^{-1}}$. For the NRO 45m COMING data, we adopt the \(\eta_{\text{mb}}\) of 0.45 at 115 GHz\footnote{https://www.nro.nao.ac.jp/~nro45mrt/html/prop/eff/eff2015.html}, and the $S/T_{\rm mb}$ calculated for $^{12}$CO(1--0) is 2.12\footnote{https://www.nro.nao.ac.jp/~nro45mrt/html/prop/plan.html}.

	The calculation of line luminosity $L^\prime_{\rm gas}$ is conducted using the equation provided in \cite{solomon_1997}:
	\begin{equation}
		\begin{aligned}
			L^\prime_{\text{gas}} &= 3.25 \times 10^7 \left(\frac{S\Delta v}{{1\ \text{Jy km s}^{-1}}}\right)
			\left(\frac{\nu_{\text{obs}}}{\text{1 GHz}}\right)^{-2} \\
			& \times\left(\frac{D_{\text{L}}}{\text{1\ Mpc}}\right)^2 \left(1+z\right)^{-3}\ [\text{K km s}^{-1} \text{pc}^2].
		\end{aligned}
	\end{equation}

	\begin{figure}[H]
		\centering
		\includegraphics[width=\hsize]{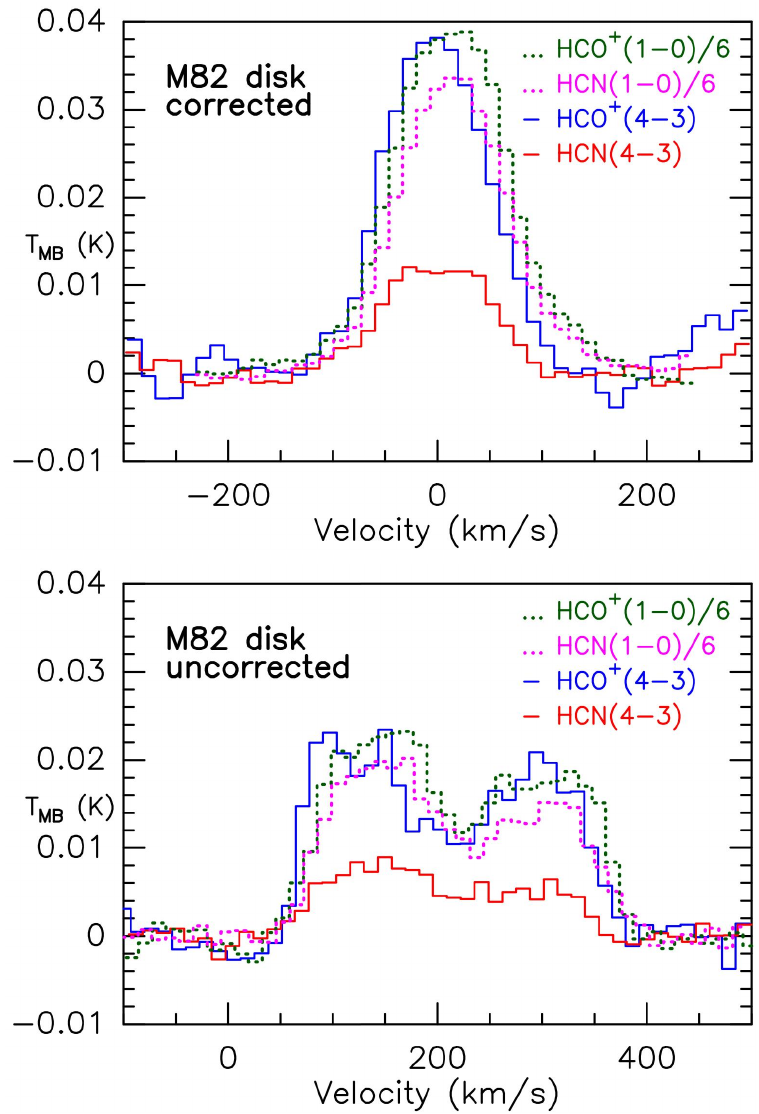}
		\caption{(\textit{top panel): } RMS-weighted spectral stacking results for the disk of M82. The velocities of dense-gas tracers were corrected according to the line center of CO(3--2) before stacking. A scaling factor of 6 was applied to HCN(1--0) and HCO$^+$(1--0). The velocities at the centers of the Gaussian fits for HCN(1--0), HCO$^+$(1--0), HCN(4--3), and HCO$^+$(4--3) are 14.3~$\pm$~0.5, 13.6~$\pm$~0.6, –0.6~$\pm$~2.5, and –1.8~$\pm$~1.9~~~km s$^{-1}$, respectively.
		(\textit{bottom panel}): RMS-weighted spectral stacking results for the disk of M82 without velocity correction.
		}
		\label{fig: stacking_disk}
	\end{figure}

	\section{Results}
	\label{sect:results}
	
	\subsection{Spectra and stacking results}
	\label{sec:Spectra}

	Figures \ref{fig: sp_disk} to \ref{fig: south_se} show the spectra  of all gas tracers.
	Given that the intensities of HCN(4--3) and HCO$^+$(4--3) are significantly lower than those of other lines, we applied scaling factors to CO(1--0), CO(3--2), HCN(1--0), and HCO$^+$(1--0) for clearer comparison.
	In most regions, different tracers  exhibit similar spatially integrated line centers and line widths. However, in certain areas, such as on the southwest (SW) side of the major axis at offset ($-$10,0) and ($-$20,0), the line center of CO(1--0)  deviates by $\sim$ 50 km\ s$^{-1}$.
	
\begin{figure*}[h]
		\centering
		\includegraphics[width=\hsize]{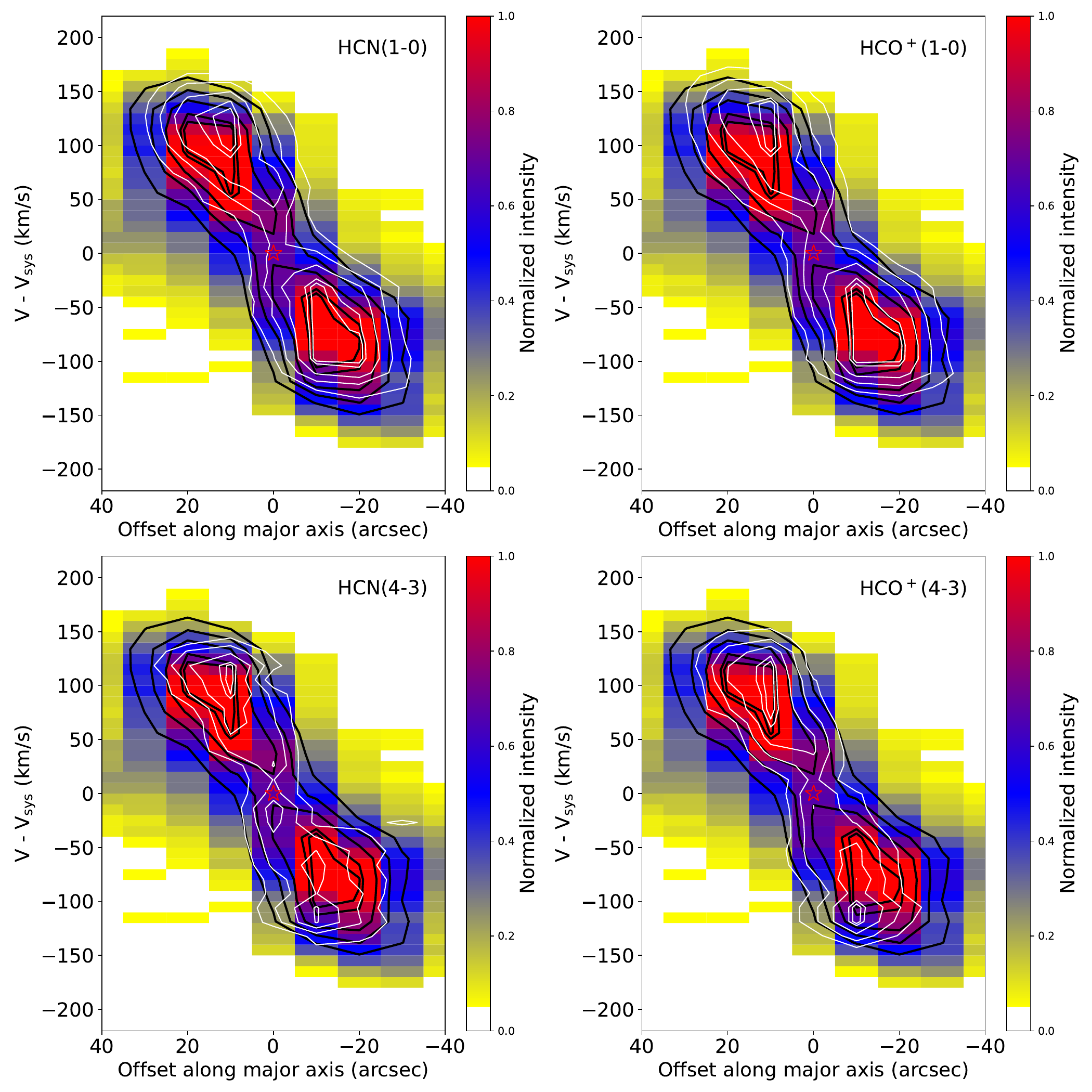}
		\caption{Position-velocity diagrams along the major axis. Velocity 0 corresponds to the systemic velocity of the galaxy, $V_{\rm sys}$(LSR) = +225 km s$^{-1}$ (\citealp{Neininger_1998}), and the position (0,0) is marked by a star. The contour levels are at 30\%, 50\%, 70\%, 90\%, and 95\% of the peak intensities. The white contours represent HCN(1--0), HCO$^+$(1--0), HCN(4--3), and HCO$^+$(4--3), while the black contours correspond to the CO(3--2). Color scale shows the normalized intensity of CO(1--0).}
		\label{fig: pv}
	\end{figure*}

	We performed spectral stacking for HCN(4--3), HCO$^{+}$(4--3), HCN(1--0), and HCO$^{+}$(1--0) in the disk regions. We calculated the weights using the RMS of HCO$^+$(1--0): weight = 1 / $\sigma^2$, and the same weight distribution was applied to all dense-gas tracers. 
	The global rotation  and local dynamics of M82 cause velocity shifts in emission lines at different regions. To achieve a higher S/N in the stacked results, we performed velocity corrections before stacking. We assumed that the velocities of dense-gas signals potentially hidden in the noise are the same as or close to the CO(3--2) velocities at the same regions. Under this assumption, we shifted the velocity centers of dense-gas lines to 0 km s$^{-1}$ based on Gaussian fits to the CO(3--2) line. We also conducted spectral stacking without velocity corrections to preserve the local dynamical characteristics across different regions.
	Figure \ref{fig: stacking_disk} presents both the velocity-corrected (top panel) and uncorrected (bottom panel) stacking results.
	In the velocity-corrected result, the velocity-integrated intensity of HCN(4--3) is lower than HCO$^+$(4--3) with a lower ratio compared to that of HCN/HCO$^+$(1--0) ratio, indicating that the excitation of high-$J$ HCN emission is significantly suppressed within M82.
	The spectral profiles of HCN(4--3) and HCO$^+$(4--3) are also more similar to each other,  those of the $J$ = 1--0 transitions. 
    However, slight shifts in peak velocity are observed between the \Jt{1}{0} and \Jt{4}{3} lines.  
	In the uncorrected result, these velocity offsets mainly appear on the blueshifted side, corresponding to the SW lobe shown in Fig.\ref{fig: pv}.

	Additionally, we performed spectral stacking for two specified regions to  search for potential \Jt{4}{3} signals.
	Region A includes areas where $\rm HCO^{+}$(1--0) was detected but $\rm HCO^{+}$(4--3) was not, while Region B includes areas where HCN(1--0) was detected but HCN(4--3) was not. The results of stacked spectra, as well as the spatial definitions of Region A and B, are presented in Fig. \ref{fig: stacking_ab}. The integrated velocity flux densities obtained from the stacking are listed in the last columns of Table \ref{tab:A2}.
	Following stacking, HCO$^+$(4--3) emission with S/N $\geq$ 3 was detected in  both regions, whereas HCN(4--3) emission with S/N $\geq$ 3 was detected only  in Region B.

	\begin{figure*}[htbp]
		\centering
		\includegraphics[width=\hsize]{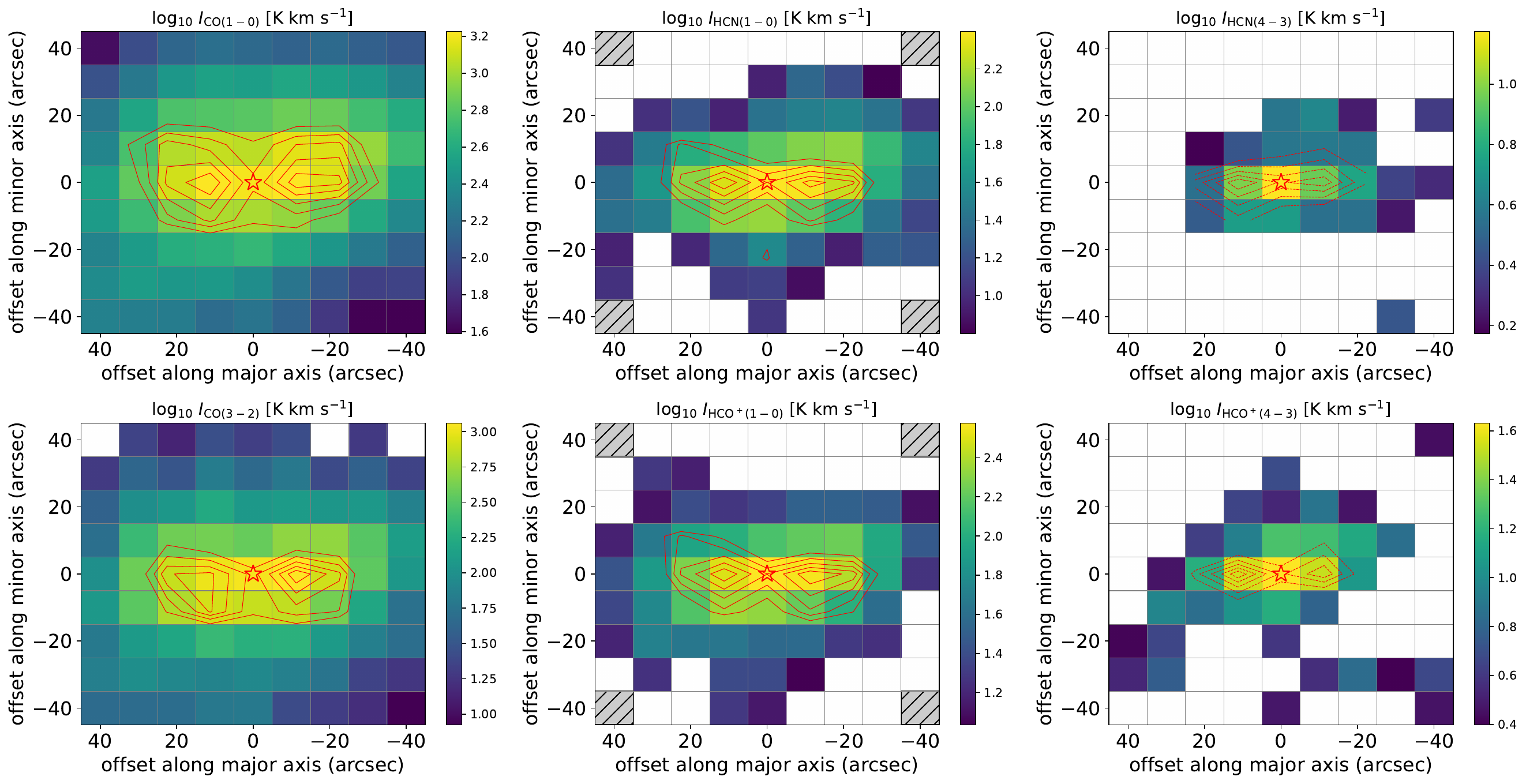}
		\caption{Maps of the entire area observed of various molecular-line emissions, color-coded by the velocity-integrated line intensity. Regions with the velocity-integrated line intensity less than 3$\sigma$ are marked by blank pixels, and the typical RMS is represented by the lowest values of each map. Each panel in the figure corresponds to a different molecular line, as indicated by the title. The galaxy's center is marked by a red star and the gray squares denote positions with no available observational data. Contours are drawn at 50\%, 60\%, 70\%, 80\%, and 90\% of the maximum of the peak main-beam temperature ($T_\mathrm{peak}$) for each line.}
		
		\label{fig: heatmap}
	\end{figure*}

\subsection{Velocity distribution}
	\label{sec:pv}
	In Fig. \ref{fig: pv},  we present the position-velocity ($p$--$v$) diagram along the major axis. The systemic velocity of M82, $V_{\rm{sys}}(\rm{LSR})$ = +225 km\,s$^{-1}$ (\citealp{Neininger_1998}) has been subtracted, and the dynamical center is marked at ($V - V_{\rm{sys}}$ = 0, offset = 0). All tracers exhibit a characteristic "figure-of-eight" pattern, with intensity offset from the center by $\pm$10 arcsec.

	On the NE side of the dynamical center (i.e., the upper left of each panel in Fig. \ref{fig: pv}), the contour shapes and peak positions of the four dense gas tracers are broadly consistent, although HCN(4--3) appears more spatially concentrated. CO(3--2) and CO(1--0) also show good spatial and kinematic agreement. 
	In contrast, on the SW side of the dynamical center (i.e., the lower right of each subpanel in Fig. \ref{fig: pv}), the peaks and contour shapes diverge significantly between different lines. In addition to the normal behaviour following the HCN(1--0) and HCO$^+$(1--0) pattern, HCN(4--3) and HCO$^+$(4--3) exhibit an additional bright region, which results in their peak velocities being about --110 km s$^{-1}$,  offset by $\sim$40 km  s$^{-1}$ from those of HCN(1--0) and HCO$^+$(1--0).

		\begin{table}[htbp]
		\caption{The luminosity ratios at NE and SW lobes}
		\centering
		\resizebox{0.49\textwidth}{!}{
			\begin{tabular}{lcc}
				\hline
				Ratios & NE peak (10,0) & SW peak (--10,0) \\
				\hline
				HCN(4--3) / HCN(1--0)   & $0.17 \pm 0.01$   & $0.14 \pm 0.01$   \\
				HCO$^+$(4--3) / HCO$^+$(1--0)   & $0.43 \pm 0.02$    & $0.36 \pm 0.01$    \\
				HCN(1--0)/CO(1--0)     & $0.11 \pm 0.01$    & $0.11 \pm 0.01$    \\
				HCO$^+$(1--0)/CO(1--0)  & $0.47 \pm 0.01$    & $0.46 \pm 0.01$    \\
				HCN(4--3) / HCO$^+$(4--3)    & $0.27 \pm 0.02$    & $0.26 \pm 0.02$    \\
				HCN(1--0) / HCO$^+$(1--0)     & $0.66 \pm 0.02$     & $0.68 \pm 0.02$    \\
				CO(3--2) / CO(1--0)   & $1.10 \pm 0.02$     & $1.20 \pm 0.01$     \\
				\hline
			\end{tabular}
		}
		
		\label{tab:ratios}
	\end{table}

	\subsection{Radial distribution}

	Figure \ref{fig: heatmap} shows the velocity-integrated intensity maps of CO(1--0), CO(3--2), HCN(1--0), HCO$^{+}$(1--0), HCN(4--3), and HCO$^{+}$(4--3) emission. The pixels with S/N $\geq$ 3 are highlighted, while  lower-significance pixels are marked as hatched pixels.
	Contours based on peak brightness temperature reveal an obvious double-peaked structure in all tracers.  
	Table \ref{tab:ratios}  summarizes the line ratios measured at the two peaks. Both the HCN(4--3)/HCN(1--0) and HCO$^{+}$(4--3)/HCO$^{+}$(1--0) ratios are higher at the NE peak, whereas the remaining ratios are comparable between the NE and SW peaks.
	
\begin{figure}[]
		\centering
		\includegraphics[width=\hsize]{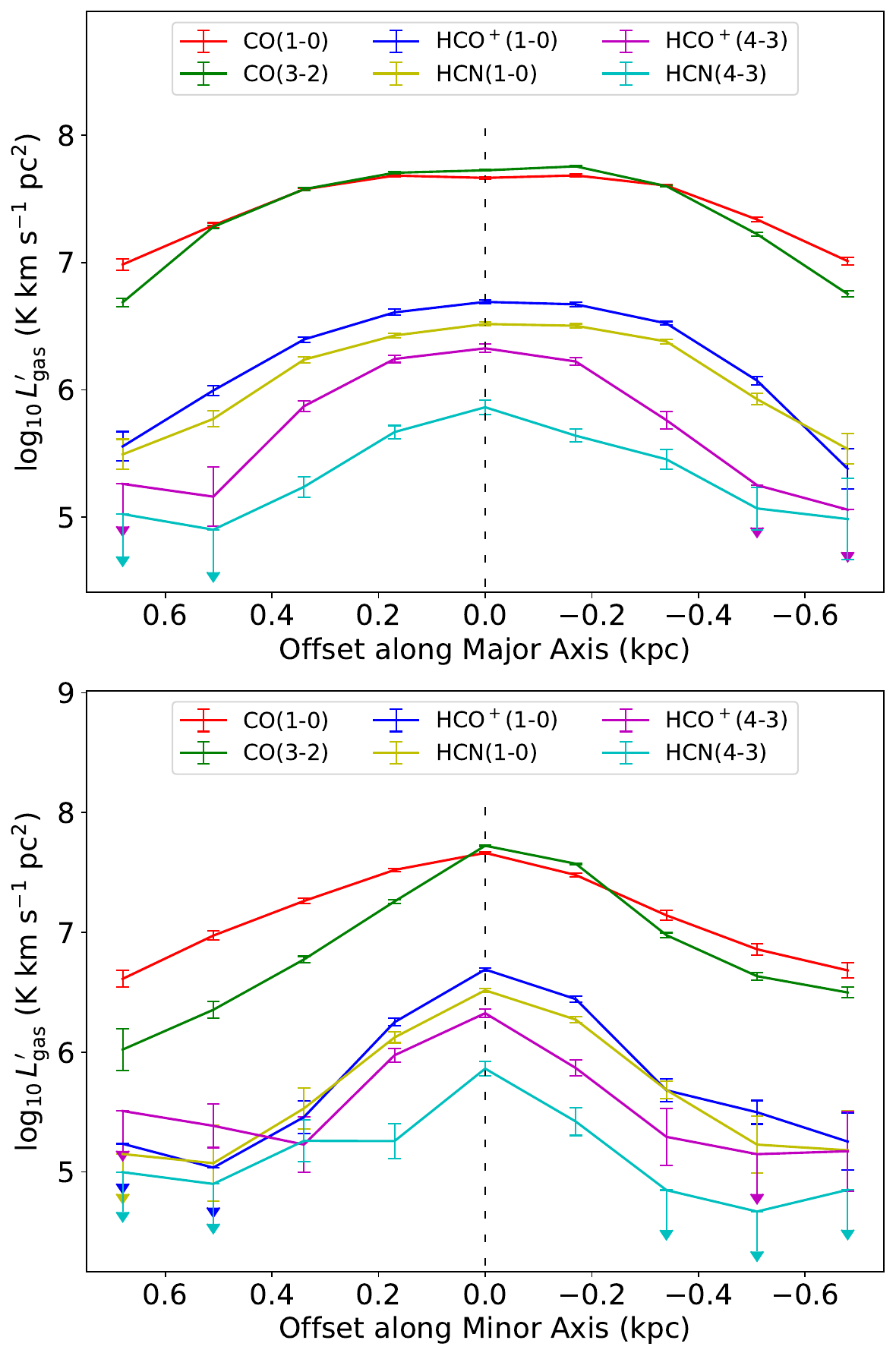}
		\caption{(\textit{top panel}): Distribution of the line luminosity of various molecular gas tracers along the major axis. A dashed line is drawn to mark offset = 0. (\textit{bottom panel}): Similar to the upper figure, but along the minor axis. Given that the inclination angle of the outflow is quite small ($\sim$ $10^\circ$, \citealp{Leroy_2015}), we assume that $1\arcsec$ along the minor axis associated with the outflow corresponds  to a projected distance of about 17 pc, the same as along the major axis.}
		\label{fig: radial}
	\end{figure}

\begin{figure}[]
		\centering
		\includegraphics[width=\hsize]{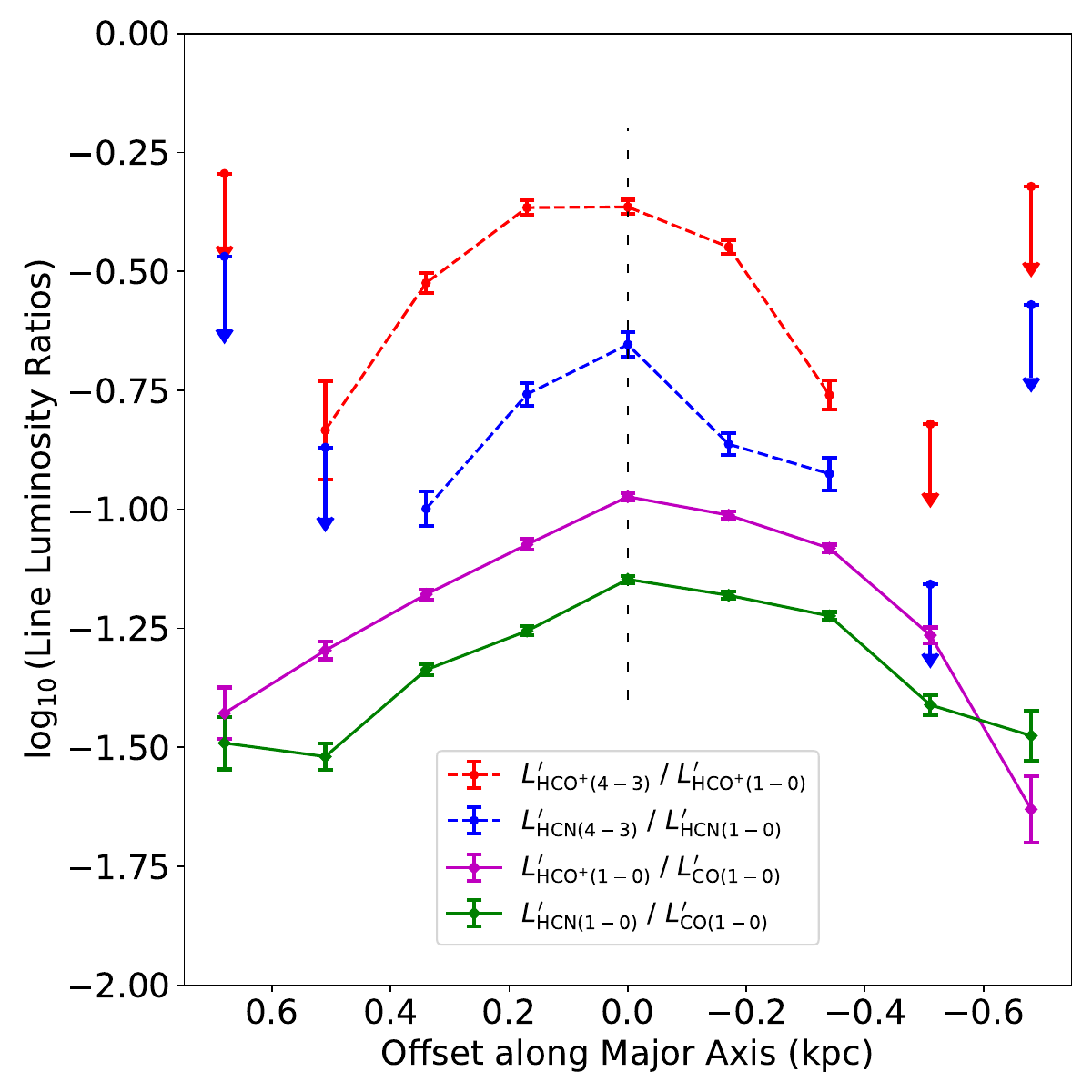}
		\caption{The luminosity ratios of HCO$^+$(4--3)/(1--0) and HCN(4--3)/(1--0) are represented by the dashed lines, and the luminosity ratios of HCN(1--0)/CO(1--0) and HCO$^+$(1--0)/CO(1--0) are represented by the solid lines. Both are measured along the major axis of M82.}
		\label{fig: Ratios_vs_distance}
	\end{figure}

	The CO(3--2)/CO(1--0) luminosity ratio is greater than unity near the galaxy's center, but declines below unity at larger radii,  consistent with the result of \citet{Salak_2013}.  This radial variation is clearly illustrated in Fig. \ref{fig: radial}, which displays the line luminosities profiles of molecular gas along both the major and minor axes.  The profiles also reveal a steeper decline along the minor axis than along the major axis.
	The luminosity ratios HCN(4--3)/(1--0), HCO$^+$(4--3)/(1--0), HCO$^+$(1--0)/CO(1--0), and HCN(1--0)/CO(1--0) all exhibit  decreasing trends with increasing projected galactic radius (Fig. \ref{fig: Ratios_vs_distance}). However, the decreasing trend differs within a radius of $\sim$400 pc. The HCN(4--3)/(1--0) and HCO$^+$(4--3)/(1--0) ratios decrease more rapidly on the  SW side, while HCO$^+$(1--0)/CO(1--0) and HCN(1--0)/CO(1--0), decrease more gradually on the in the same direction.

\begin{figure*}[htbp]
		\centering
		\includegraphics[width=\hsize]{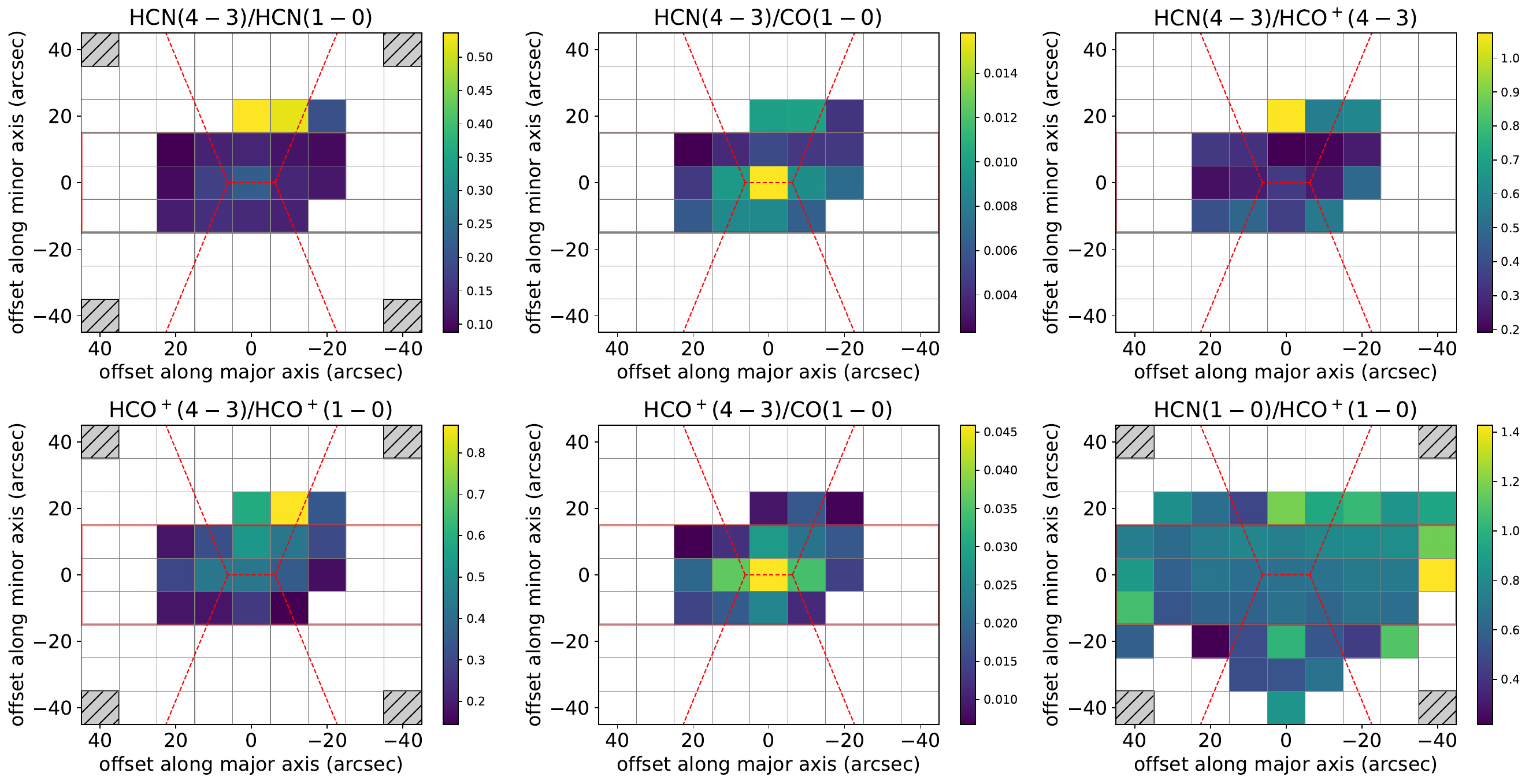}
		\caption{Distribution of the luminosity ratios HCN (4--3)/(1--0), $ \rm HCO^{+}$(4--3)/(1--0), HCN(4--3)/CO(1--0), HCO$^+$(4--3)/CO(1--0), HCN/$ \rm HCO^{+}$(4--3) and HCN/$ \rm HCO^{+}$(1--0). Here, only regions with velocity-integrated line intensities exceeding 3$\sigma$ for both lines are included, with the remaining pixels shown as blank. For the maps involving HCN(4--3) or HCO$^+$(4--3), we applied a selection where both HCN(4--3) and HCO$^+$(4--3) have S/N $\geq$ 3, with the remaining points marked as hatched. For the HCN/HCO$^+$(1--0) map, we used a similar S/N $\geq$ 3 filter for both HCN(1--0) and HCO$^+$(1--0). Similar to Fig. \ref{fig: position}, we use red rectangles to indicate the disk-dominated region and a cone with a base of 300 pc and an opening angle of 20$^\circ$ to represent the regions associated with outflow.}
		\label{fig: heatmap_raiots}
	\end{figure*}

	The luminosity ratios HCN(4--3)/(1--0), HCO$^+$(4--3)/(1--0), HCN(4--3)/CO(1--0), HCO$^+$(4--3)/CO(1--0), HCN/HCO$^+$(4--3), and HCN/HCO$^+$(1--0) across all observed positions are shown in Fig. \ref{fig: heatmap_raiots}.
	Among these, only the HCN(4--3)/CO(1--0) and HCO$^+$(4--3)/CO(1--0) ratios peak at the galaxy's center. Interestingly, the HCN(4--3)/(1--0) and HCO$^+$(4--3)/(1--0) ratios peak in the northwest side region, where strong multiphase outflows  have been detected (e.g., \citealp{Leroy_2015}). 
	Besides, the HCN/HCO$^+$(4--3) and HCN/HCO$^+$(1--0) ratios also show relatively high values  in the same area.
	In all regions where \Jt{4}{3} was detected, the HCN(4--3)/(1--0) ratio is consistently lower than HCO$^+$(4--3)/(1--0), and HCN/HCO$^+$(4--3) is also lower than HCN/HCO$^+$(1--0) (see Table \ref{tab:Luminosity Ratios}).

\section{Analysis and discussion}

	\label{sect:discussion}
	
	\subsection{Line ratios}
	
		Along the galaxy's major axis,  the line ratios between different transitions of the same molecule clearly decrease with increasing distance from the center. 
		Figure \ref{fig: Ratios_vs_distance} shows a significant decline in the HCO$^+$(4--3)/(1--0) ratio around 0.2--0.5 kpc SW side of the center, while the HCN(4--3)/(1--0) ratio remains consistently low along the major axis. 
		The HCN(4--3)/(1--0) ratio reaches a peak of $\sim$0.2 at the center and decreases to $\sim$0.1 at an offset of 20\arcsec.  In contrast, the HCO$^+$(4--3)/(1--0) ratio declines from $\sim$0.4 at the center to about 0.2--0.3 (see Table \ref{tab:Luminosity Ratios}). 
		Compared with the average values of $0.27 \pm 0.04$ for HCN(4--3)/(1--0) and $0.29 \pm 0.07$ for HCO$^+$(4--3)/(1--0), derived from 22\arcsec-beam observations toward the centers of galaxies including starbursts, AGNs, and LIRG/ULIRG \citep{Isreal_2023}, M82 exhibits a lower HCN(4--3)/(1--0) ratio and a higher HCO$^+$(4--3)/(1--0) ratio.
		
		Also, we calculated the ratios of HCN(4--3) to $ \rm HCO^{+}$(4--3) where the intensity was greater than 3$\sigma$ (see Table \ref{tab:Luminosity Ratios}), finding them to range from $0.11 \pm 0.02$ to $0.79 \pm 0.16$,  with an average value of $0.36 \pm 0.06$.
		The mean value of HCN/HCO$^+$(1--0) across all detected locations is 0.66 $\pm$ 0.09, and 0.70 $\pm$ 0.04 in the central area where  the $J$ = 4--3 transition was detected in both species. In comparison, the average intensity ratio of HCN/HCO$^+$(1--0) of the nine EMPIRE galaxies in \cite{Jim_2019}  is 1.43 $\pm$ 0.41, and in the 43 nearby galaxies studied by \cite{Isreal_2023}, this ratio of HCN/HCO$^+$(1--0) is 1.11 $\pm$ 0.06. 
		Meanwhile, \cite{Isreal_2023} also reports an average HCN/HCO$^+$(4--3) ratio of 0.70 $\pm$ 0.10 for 15 galaxies with HCN(4--3) and HCO$^+$(4--3) detections.
		Both the HCN/HCO$^+$(1--0) and HCN/HCO$^+$(4--3) line ratios of M82 are lower than the average values reported in the literature. 
	
		The spatial distributions of HCN/HCO$^+$  in the NE and SW lobes are comparable, with no significant asymmetry. In the central starburst region (within $\sim$0.5~kpc), both sides show similar HCN/HCO$^+$(1--0) values of $\sim$0.65--0.70. At higher excitation, the HCN/HCO$^+$(4--3) ratio remains low ($\sim$0.25--0.30). 
		Although the ratios are generally uniform, localized departures are observed in specific regions.
		For example, around 0.2--0.5 kpc SW side of the center, the HCO$^+$(4--3) intensity drops significantly (see Fig.~\ref{fig: radial}), leading to a slight increase in the HCN/HCO$^+$(4--3) ratio there. However, the ratio remains low overall, suggesting that HCO$^+$ emission dominates over HCN, especially at higher-$J$ transitions.

		It has been suggested that supernova explosions and photo-dissociation regions (PDRs) can significantly increase the abundance of HCO$^+$ and decrease the abundance of HCN (e.g., \citealt{Wild_1992}; \citealt{Schilke_2001}), which is consistent with the pure-starburst property of M82. The gas disk of M82 contains a giant PDR that is 650~pc in size, which may significantly increase the abundance of HCO$^+$ in M82 (e.g., \citealt{Garcia-Burillo_2002}; \citealt{Fuente_2005}; \citealt{Krips_2008}).

		Theoretical models also show that in PDRs with densities below $10^5$~cm$^{-3}$, the HCN/HCO$^+$ ratio drops below unity (e.g., \citealp{Meijerink_2007, Yamada_2007}). Multi-component PDR models by \citet{Mao_2000} and \citet{loenen_2010} suggest that most molecular gas in M82 is in low-excitation diffuse components with densities around $n$(H$_2$) = $10^3$ -- $10^{3.7}$~cm$^{-3}$, and only a small fraction ($\sim$1\%) is in dense gas with $n$(H$_2$) = $10^6$~cm$^{-3}$.
		Under optically thin conditions and at 20~K, the critical densities ($n_{\rm crit}$) of HCO$^+$(1--0), HCN(1--0), HCO$^+$(4--3), and HCN(4--3) are $4.5 \times 10^4$, $3.0 \times 10^5$, $3.2 \times 10^6$, and $2.3 \times 10^7$~cm$^{-3}$, respectively \citep{Shirley_2015}. 
		Given its higher critical density, HCN(4--3) is likely not fully excited in M82's relatively low average-density environment, while the excitation of HCO$^+$ may be maintained at a higher level under the influence of PDRs. The combined effect of these two factors may contribute to the relatively low HCN/HCO$^+$ ratio observed in M82 and its further decrease with increasing rotational transition.

	\subsection{Comparison with previous measurements}

Our data reveal that the molecular emission in M82 exhibits a characteristic double-lobed structure, with prominent peaks located to the NE and SW of the nucleus, consistent with previous CO observations (e.g., \citealt{Loiseau_1990,Kikumoto_1998,Neininger_1998,Mao_2000,Wei_2001,walter_2002,Ward_2003,Seaquist_2006,loenen_2010}).
As shown in Fig. \ref{fig: sp_disk}, the spectral line widths at the central position (0,0) are broader than those at the NE lobe (10,0) and the SW lobe ($-$10,0), while the peak line temperatures at the center are lower than those in the lobes. This trend is also evident in Fig. \ref{fig: heatmap}: although the velocity-integrated intensities reach a maximum at the (0,0) position, the contour maps of $T_{\rm peak}$ show a double-peaked structure. This is likely due to the central beam encompassing partial emission from both lobes, resulting in spatial blending that broadens the line profile and reduces the observed peak brightness temperature.

Table~\ref{tab:intensity} summarizes the integrated intensities of various emission lines measured at the NE and SW lobes, along with their corresponding SW-to-NE intensity ratios. In terms of absolute values, the CO line intensities decrease with increasing rotational quantum number in both lobes, with a steeper decline observed from CO(5--4) onward. It is important to note that these data were obtained with different telescopes and beam sizes, therefore intensity ratios between different transitions may not directly reflect physical conditions. Comparisons of the NE and SW lobes with the same observation are more reliable. For most molecular lines, the SW lobe shows slightly stronger emission than the NE lobe, with SW/NE intensity ratios typically ranging from 1.1 to 1.3.
A similar trend is observed in the $^{13}$CO lines, where transitions from \Jt{1}{0} to 5$-$4 exhibit SW/NE ratios of about 1.2 to 1.5, consistent with those of the CO results. An extreme case is $^{13}$CO(6--5), which shows a SW/NE intensity ratio of $\sim$2.8 $\pm$ 0.88, likely due to the very low intensity at the NE lobe (\citealp{loenen_2010}). 
In contrast, our HCN(4--3) and HCO$^+$(4--3) lines, along with CO(10--9) from \cite{loenen_2010}, show SW/NE ratios below unity. This may suggest an enhanced fraction of dense gas toward the NE lobe.

Similar to the results shown in Fig. \ref{fig: pv}, previous studies have also reported a prominent double-lobed structure in the $p$--$v$ diagrams of various emission lines. The two lobes are roughly symmetric in spatial position, located about 10-15 arcsec on either side of the center, but their velocity distributions show subtle differences. The NE lobe exhibits consistent velocities across different lines, with no clear evidence of velocity stratification. 
In contrast, the SW lobe shows a truncation in velocity: the velocities of low- to mid-$J$ CO lines ($J\le4$) and low-$J$ ($J=1$) HCN (HCO$^+$) lines are concentrated around 140-170 km s$^{-1}$ (e.g., \citealt{Mao_2000,Ward_2003}), while the peak velocities of high-$J$ CO and high-$J$ HCN and HCO$^+$ transitions are closer to $\sim$110 km s$^{-1}$ (e.g., \citealt{loenen_2010,Ward_2003,Mao_2000,Wild_1992}). This suggests that the SW lobe contains multiple gas components with distinct excitation conditions and kinematic properties.

\citet{Mao_2000} reported that the spatial separation between the NE and SW emission peaks in their velocity-integrated intensity maps decreases significantly with increasing rotational quantum number $J$: from 26{\textrm \arcsec} in CO(1--0) and CO(2-1) to 15{\textrm \arcsec} in CO(7--6). They argued that this difference exceeds the beam size and positional uncertainties, and is therefore unlikely to be caused by resolution effects. However, \citet{Mao_2000} noted that this truncation is not clearly apparent in $p$--$v$ diagrams. By incorporating additional emission lines, particularly our new HCN(4--3) and HCO$^+$(4--3) observations, we confirm the presence of excitation-dependent velocity truncation. Interestingly, our results reveal an opposite trend in the $p$--$v$ diagrams compared to the velocity-integrated maps: on the SW side, high-excitation lines peak at lower velocities, resulting in a larger velocity separation between the NE and SW lobes. This feature is also clearly seen in Fig. 4 of \citet{Ward_2003}.
In addition, \cite{loenen_2010} reported a secondary velocity component near 100 km s$^{-1}$ in the SW region using Herschel high-$J$ CO spectra. In the CO \Jt{9}{8} and higher transitions,  this secondary component is even stronger than the main peak around 160 km s$^{-1}$,  with the intensities scaled to a common beam size. This result is consistent with the lower-velocity emission we observe in high-excitation lines on the SW side.
These findings suggest that while high-excitation lines are more spatially concentrated toward the galaxy's center, the observed velocity 
offset in the SW lobe may reflect a localized dynamical feature, possibly associated with non-circular motions or feedback-driven structures that deviate from the overall disk rotation.

	\section{Summary} 
	\label{sect:conc}
	
	We have presented $ \rm HCO^{+}$ \Jt{4}{3} and HCN \Jt{4}{3} observations of the central $90\arcsec~\times~90\arcsec$  region of M82 obtained as part of the JCMT MALATANG program. By combining these data with the HCO$^{+}$~\Jt{1}{0} and HCN \Jt{1}{0} 
	data from the GBT, as well as  CO data from the JCMT NGLS survey and Nobeyama 45m COMING project, we investigated the gas properties on sub-kpc scales.  The main results and conclusions of this study are summarized as follows:
	
	\begin{enumerate}

		\item All of the emission lines analyzed, CO(1--0), CO(3--2), HCN(1--0), HCO$^+$(1--0), HCN(4--3), and HCO$^+$(4--3) peak near the center of M82 and  exhibit declining intensities with increasing distance from the center along both the major and minor  axes of the galaxy. In M82, the emission from HCN(4--3) and HCO$^+$(4--3) is primarily concentrated within $\sim$500 pc of the galaxy's center. To probe low-level emission in two fainter regions, we applied spectral stacking, which revealed a weak HCO$^+$(4--3) signal and provided an upper limit on the velocity-integrated flux density of HCN(4--3).

		\item The HCN(4--3)/(1--0) line ratio ranges from 0.09 to 0.54, with a mean value of 0.18 $\pm$ 0.04, while the HCO$^+$(4--3)/(1--0) line ratio ranges from 0.14 to 0.87, with  a mean of 0.32 $\pm$ 0.06.  For both tracers, the highest ratios are found on the northwest side of the galaxy, spatially coinciding with M82's well-known outflow region. The low-$J$ transitions, \Jt{1}{0} of HCN and HCO$^+$, are brighter on the SW side of the major axis, while the high-$J$ transitions, \Jt{4}{3} of HCN and HCO$^+$ are stronger on the NE side.  In all regions where \Jt{4}{3} transitions are detected, the HCN(4--3)/(1--0) ratio is consistently lower than the corresponding HCO$^+$(4--3)/(1--0) ratio. The HCN/HCO$^+$(4--3) ratio ranges from 0.19 $\pm$ 0.03 to 1.07 $\pm$ 0.31, with a mean value of 0.39 $\pm$ 0.11, which is lower than the average HCN/HCO$^+$ (1--0) ratio of 0.74 $\pm$ 0.08. This difference may be attributed to the relatively low excitation of high-$J$ HCN transitions.

		\item 	The position-velocity diagram along the major axis reveals the characteristic "figure-of-eight" pattern of M82's rotating molecular disk, with the NE and SW lobes symmetrically distributed on either side of the nucleus. In the NE lobe, the velocities traced by different lines are closely aligned, indicating a coherent kinematic structure. In contrast, the SW lobe exhibits a $\sim$40 km s$^{-1}$ blueshift in the peak velocities of high-excitation lines relative to those of the low-$J$ transitions. This velocity offset may indicate the presence of gas components with distinct excitation and kinematic properties in the SW region.

	\end{enumerate}

	\begin{acknowledgements}
		This paper is dedicated to the memory of late Professor Yu Gao, co-PI of the MALATANG project, R.I.P.
		We thank the anonymous referee for the constructive and helpful suggestions that improved the paper.
		This work was supported by National Natural Science Foundation of China (NSFC) grant No. 12033004 and National Key R\&D Program of China grant  No. 2017YFA0402704.
		We thank Pedro Salas for kindly providing the GBT data.
		This publication made use of data from COMING, CO Multi-line Imaging of Nearby Galaxies, a legacy project of the Nobeyama 45-m radio telescope. The Nobeyama 45-m radio telescope is operated by Nobeyama Radio Observatory, a branch of National Astronomical Observatory of Japan.
		We are very grateful to the EAO/JCMT staff for their help during the observations and data reduction.
		The James Clerk Maxwell Telescope is operated by the East Asian Observatory on behalf of The National Astronomical Observatory of Japan; Academia Sinica Institute of Astronomy and Astrophysics; the Korea Astronomy and Space Science Institute; the National Astronomical Research Institute of Thailand; Center for Astronomical Mega-Science (as well as the National Key R\&D Program of China with No. 2017YFA0402700). Additional funding support is provided by the Science and Technology Facilities Council of the United Kingdom and participating universities and organizations in the United Kingdom and Canada. The authors wish to recognize and acknowledge the very significant cultural role and reverence that the summit of Maunakea has always had within the indigenous Hawaiian community.  We are most fortunate to have the opportunity to conduct observations from this mountain.
		MALATANG is a JCMT Large Program with project code M16AL007 and M20AL022. We are grateful to P. P. Papadopoulos for his generous help and support with the JCMT observations.
		Z.Y.Z. acknowledges the support of the NSFC under grants No. 12173016, 12041305, the science research grants from the China Manned Space Project with NOs.CMS-CSST-2021-A08 and CMS-CSST-2021-A07, and the Program for Innovative Talents, Entrepreneur in Jiangsu.
		A.C. acknowledges support by the National Research Foundation of Korea (NRF), grant Nos. 2022R1A2C100298213 and 2022R1A6A1A03053472. 
		B.L. acknowledges support by the National Research Foundation of Korea (NRF), grant Nos. 2022R1A2C100298213.
		LCH was supported by the National Science Foundation of China (11991052, 12233001), the National Key R\&D Program of China (2022YFF0503401), and the China Manned Space Project (CMS-CSST-2021-A04, CMS-CSST-2021-A06).
		The work of MGR is supported by the international Gemini Observatory, a program of NSF NOIRLab, which is managed by the Association of Universities for Research in Astronomy (AURA) under a cooperative agreement with the U.S. National Science Foundation, on behalf of the Gemini partnership of Argentina, Brazil, Canada, Chile, the Republic of Korea, and the United States of America.
		M.J.M.~acknowledges the support of the National Science Centre, Poland through the SONATA BIS grant 2018/30/E/ST9/00208. This research was funded in part by the National Science Centre, Poland (grant number: 2023/49/B/ST9/00066).
		T.F. was supported by the National Natural Science Foundation of China under Nos. 11890692, 12133008, 12221003, and the science research grant from the China Manned Space Project with No. CMS-CSST-2021-A04.
	\end{acknowledgements}

\begin{appendix}

\onecolumn
		\section{Spectra and measurements}
		
\begin{figure}[h]
	\centering
	\includegraphics[width=\textwidth]{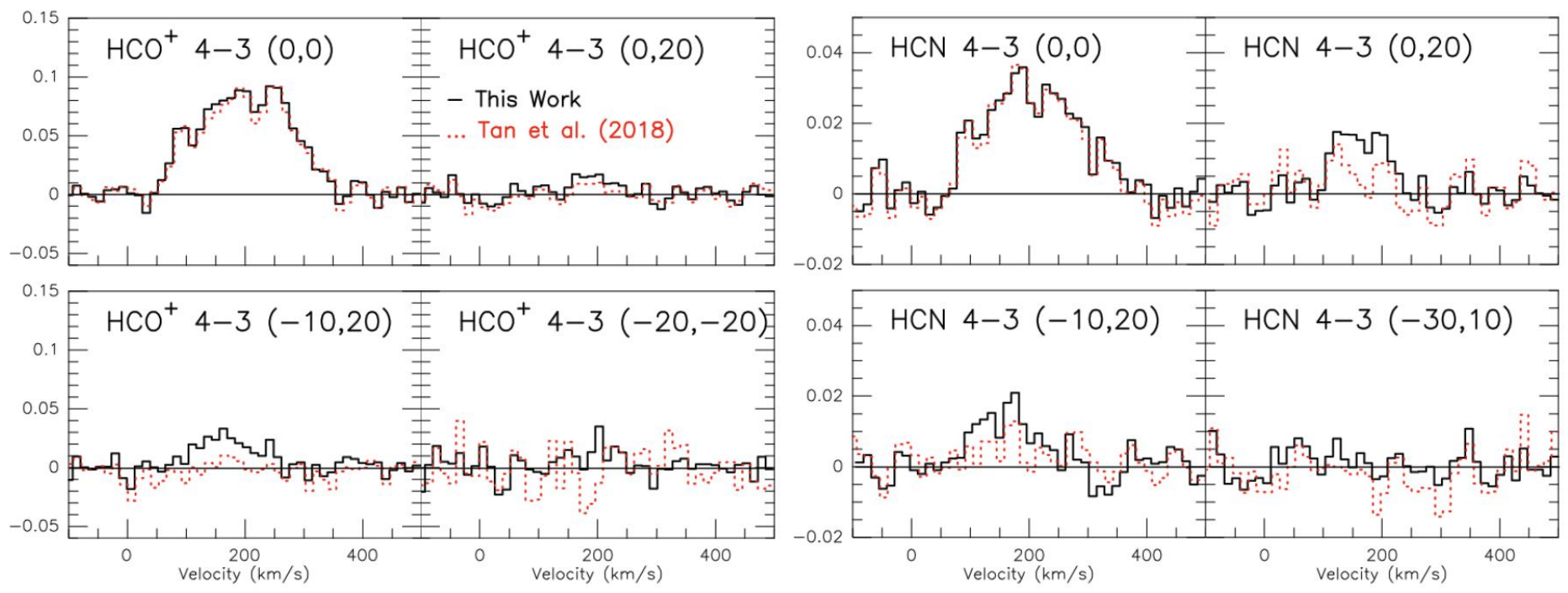}
	\caption{Selection of  spectra in units of $T_A$ (K) from this work and \cite{tan2018} are compared using solid black lines and red dotted lines, respectively. Each panel is labeled with the tracer, and the position pointed at is provided as an offset along the major and minor axes.}
	\label{fig: A1}
\end{figure}

\begin{figure}[h]
	\centering
	\includegraphics[width=0.85\textwidth]{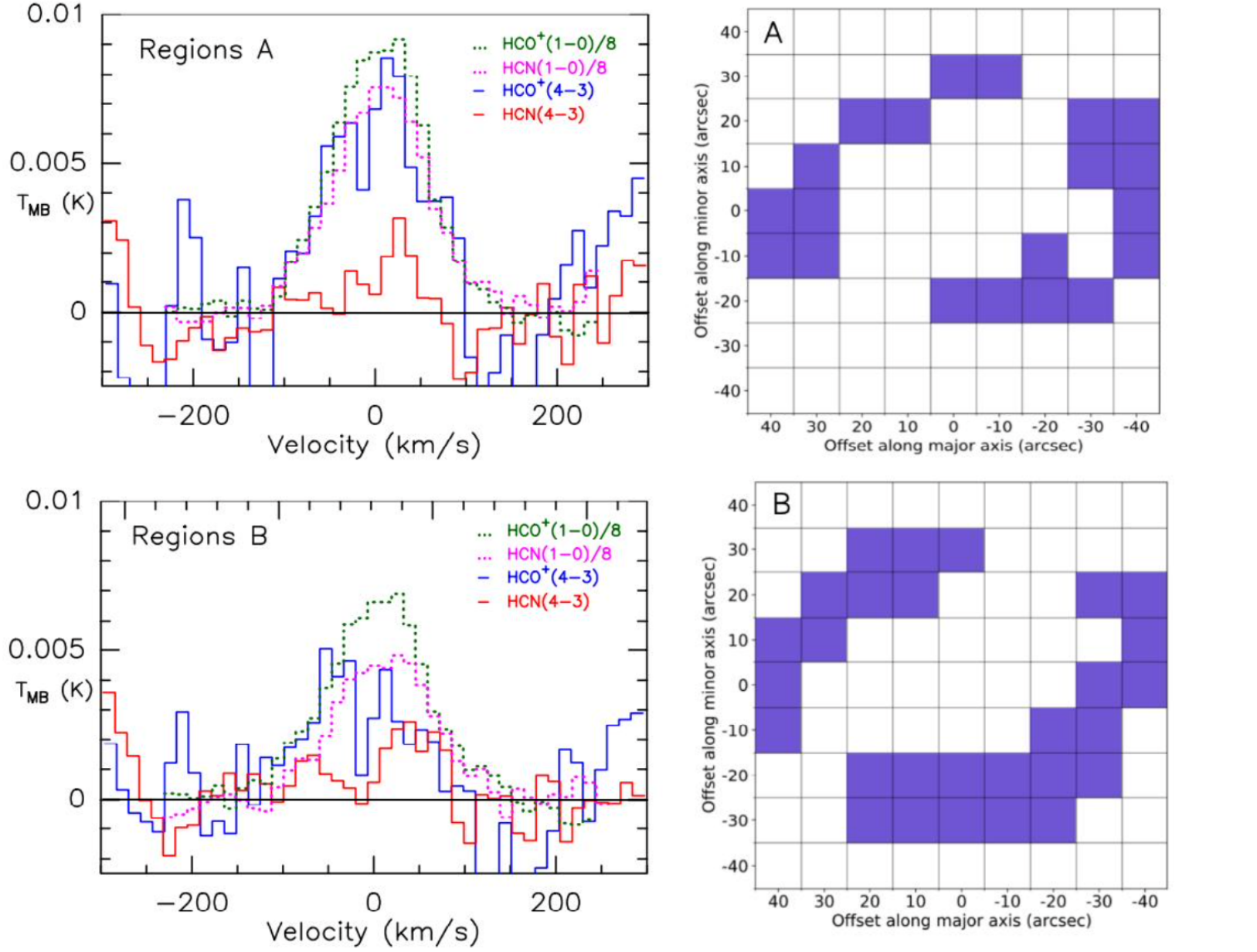}
	\caption{RMS-weighted stacked spectra for Region A and Region B are shown in the left panel. Region A represents areas where HCN(1--0) is detected but HCN(4--3) is not, while Region B corresponds to areas where $\rm HCO^{+}$(1--0) is detected but $\rm HCO^{+}$(4--3) is not. To minimize interference from high-noise spectra, data points at offsets along the minor axis of $\pm$40 have been excluded. The specific locations of Regions A and B are shown in the right panel.}
	\label{fig: stacking_ab}
\end{figure}

\clearpage

\begin{figure}[htbp]
	\centering
	\includegraphics[width=\hsize]{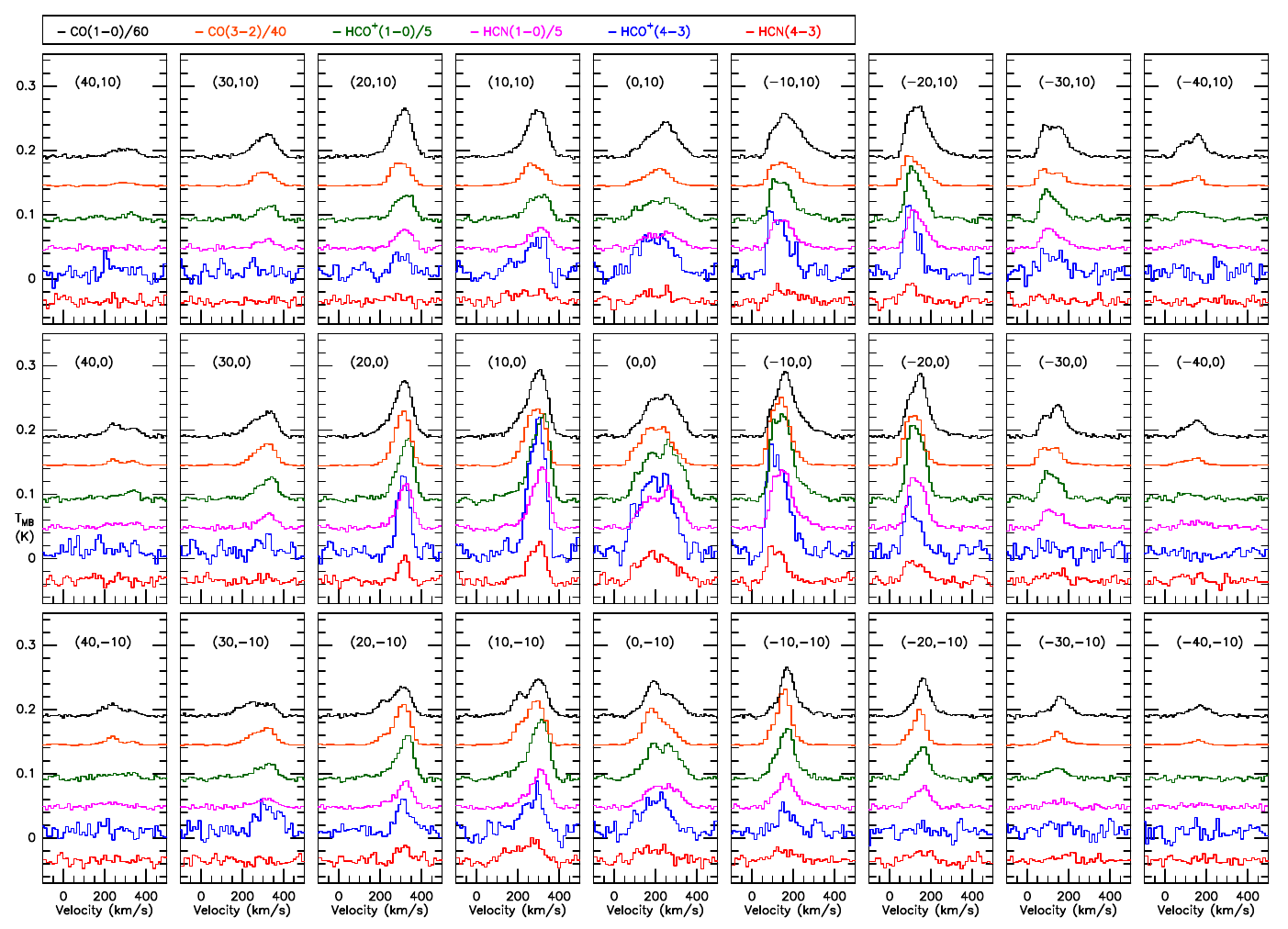}
	\caption{The spectra of CO(1--0), CO(3--2), HCO$^+$(1--0), HCN(1--0), HCO$^+$(4--3), and HCN(4--3) are shown from top to bottom in each grid, in units of $T_{\text{MB}}$ (K). Spectral are referenced to the LSR velocity reference system and the radio definition for the Doppler shift is adopted. The spectra shown here correspond exactly to the red-rectangle region in Fig. \ref{fig: position}. 
		The offset from the center position in units of arcsec is annotated in the top corner of each panel. 
		The offset (0,0) corresponds to the galaxy's center, marked by an orange circle in Fig. \ref{fig: position}. Offsets are positive toward the northeast along the major axis and toward the northwest along the minor axis, while negative offsets represent the opposite directions.
		The legend at top of figure shows the scaling factors for CO, HCN(1--0), and $ \rm HCO^{+}$(1--0). The velocity resolutions for HCN(4--3), HCO$^+$(4--3), HCN(1--0), and HCO$^+$(1--0) are $\sim$13~km\ s$^{-1}$. For CO(1--0) and CO(3--2), the velocity resolutions are 10~km\ s$^{-1}$ and 19~km\ s$^{-1}$, respectively. All data are re-sampled to a pixel size of 14 arcsec, corresponding to $\sim$240 pc.}
	\label{fig: sp_disk}
\end{figure}

\begin{figure}
	\centering
	\includegraphics[width=0.85\textwidth]{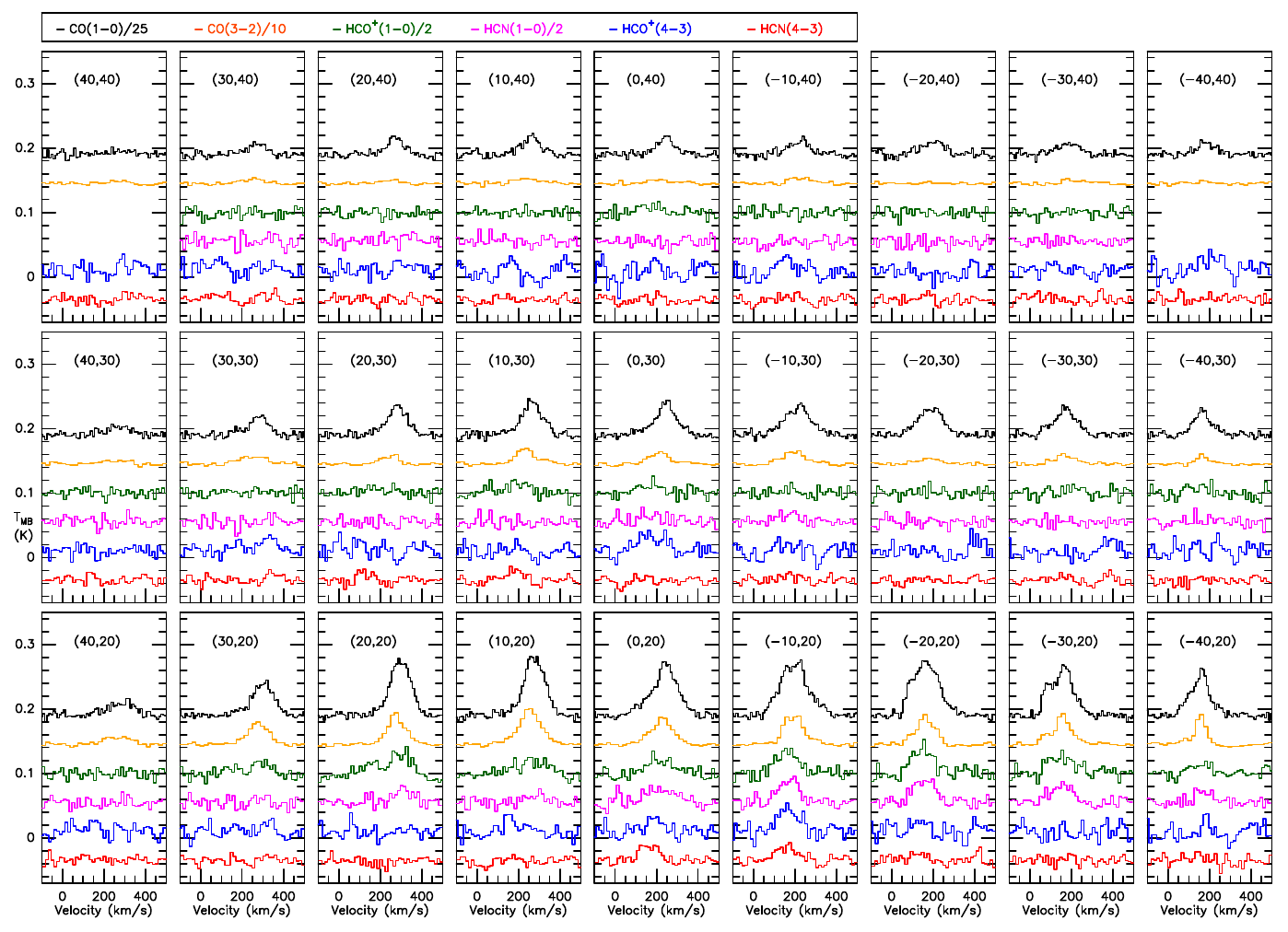}
	\caption{A continuation of Fig. \ref{fig: sp_disk}, offset along the minor axis at 20, 30, and 40 arcsec.}
	\label{fig: north_se}
\end{figure}

\begin{figure}
	\centering
	\includegraphics[width=0.85\textwidth]{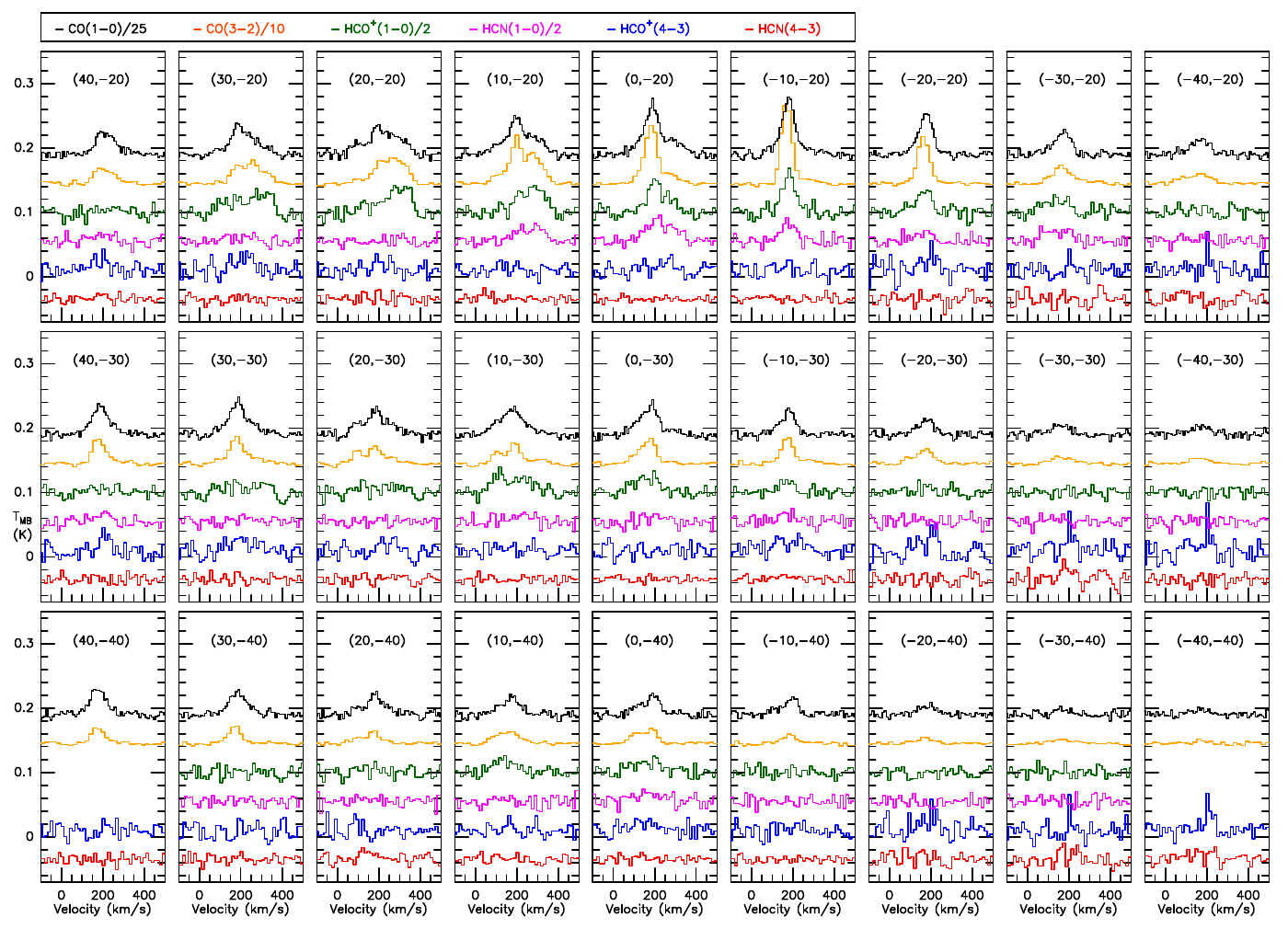}
	\caption{A continuation  of Fig. \ref{fig: sp_disk}, offset along the minor axis at $-$20, $-$30, and $-$40 arcsec.}
	\label{fig: south_se}
\end{figure}

	\begin{sidewaystable}
		\scriptsize
		\caption{Integrated intensities and luminosities}
        \label{tab:A1}
		\begin{tabular}{*{13}{>{\centering\arraybackslash}p{1.5cm}}}
			\hline
			offset & 
			$I_{\rm HCN(4--3)}$ & 
			$I_{\rm HCO^+(4--3)}$ &
			$I_{\rm HCN(1--0)}$& 
			$I_{\rm HCO^+(1--0)}$ & 
			$I_{\rm CO(1--0)}$ &  
			$I_{\rm CO(3--2)}$ &
			$L'_{\rm HCN(4--3)}$ & 
			$L'_{\rm HCO^+(4--3)}$ &
			$L'_{\rm HCN(1--0)}$& 
			$L'_{\rm HCO^+(1--0)}$ & 
			$L'_{\rm CO(1--0)}$ & 
			$L'_{\rm CO(3--2)}$ \\
			(arcsec) &  
			(${\rm K\,km\,s^{-1}}$) &  
			(${\rm K\,km\,s^{-1}}$) &  
			(${\rm K\,km\,s^{-1}}$) &  
			(${\rm K\,km\,s^{-1}}$) &  
			(${\rm K\,km\,s^{-1}}$) &  
			(${\rm K\,km\,s^{-1}}$) &   
			(${\rm 10^4\,K\,km\,s^{-1}\,pc^{2}}$) & 
			(${\rm 10^4\,K\,km\,s^{-1}\,pc^{2}}$) & 
			(${\rm 10^4\,K\,km\,s^{-1}\,pc^{2}}$) & 
			(${\rm 10^4\,K\,km\,s^{-1}\,pc^{2}}$) & 
			(${\rm 10^6\,K\,km\,s^{-1}\,pc^{2}}$) & 
			(${\rm 10^6\,K\,km\,s^{-1}\,pc^{2}}$)\\    
			\hline
			0, 0 & 9.6$\pm$0.6 & 27.4$\pm$0.9 & 65.0$\pm$0.9 & 98.1$\pm$1.3 & 724.3$\pm$6.7 & 686.0$\pm$3.4 & 72.9$\pm$4.2 & 211.6$\pm$6.9 & 328.5$\pm$4.7 & 490.0$\pm$6.4 & 46.1$\pm$0.4 & 52.9$\pm$0.3 \\
			10, 0 & 6.1$\pm$0.3 & 22.6$\pm$0.6 & 52.8$\pm$1.0 & 81.2$\pm$1.9 & 755.3$\pm$8.1 & 656.5$\pm$6.1 & 46.6$\pm$2.4 & 174.6$\pm$4.9 & 266.9$\pm$5.1 & 405.5$\pm$9.6 & 48.1$\pm$0.5 & 50.6$\pm$0.5 \\
			20, 0 & 2.3$\pm$0.2 & 9.6$\pm$0.4 & 34.1$\pm$0.8 & 49.7$\pm$1.1 & 588.5$\pm$6.3 & 487.8$\pm$5.1 & 17.3$\pm$1.4 & 74.2$\pm$3.2 & 172.3$\pm$4.2 & 248.3$\pm$5.3 & 37.5$\pm$0.4 & 37.6$\pm$0.4 \\
			30, 0 & \textless 1.0 & 1.9$\pm$0.4 & 11.7$\pm$0.7 & 19.8$\pm$0.8 & 307.1$\pm$6.1 & 246.9$\pm$3.1 & \textless 8.1 & 14.5$\pm$3.4 & 59.1$\pm$3.6 & 98.8$\pm$3.8 & 19.6$\pm$0.4 & 19.0$\pm$0.2 \\
			40, 0 & \textless 1.4 & \textless 2.4 & 6.1$\pm$0.7 & 7.2$\pm$0.8 & 151.4$\pm$6.8 & 63.1$\pm$2.0 & \textless 10.5 & \textless 18.3 & 31.1$\pm$3.7 & 35.9$\pm$4.2 & 9.6$\pm$0.4 & 4.9$\pm$0.2 \\
			--10, 0 & 5.7$\pm$0.3 & 21.6$\pm$0.6 & 62.9$\pm$1.0 & 93.9$\pm$1.6 & 757.6$\pm$7.5 & 736.9$\pm$3.2 & 43.6$\pm$2.2 & 166.8$\pm$4.9 & 318.1$\pm$4.9 & 468.8$\pm$7.8 & 48.2$\pm$0.5 & 56.8$\pm$0.3 \\
			--20, 0 & 3.7$\pm$0.3 & 7.5$\pm$0.5 & 47.4$\pm$0.8 & 66.5$\pm$1.0 & 630.2$\pm$6.4 & 515.0$\pm$2.4 & 28.4$\pm$2.2 & 57.7$\pm$4.0 & 239.4$\pm$4.0 & 332.2$\pm$4.9 & 40.1$\pm$0.4 & 39.7$\pm$0.2 \\
			--30, 0 & 1.5$\pm$0.3 & \textless 2.3 & 16.7$\pm$0.7 & 23.6$\pm$0.8 & 341.3$\pm$6.4 & 216.3$\pm$3.2 & 11.7$\pm$2.0 & \textless 17.7 & 84.3$\pm$3.8 & 118.1$\pm$4.0 & 21.7$\pm$0.4 & 16.7$\pm$0.2 \\
			--40, 0 & 1.3$\pm$0.4 & \textless 1.5 & 6.8$\pm$0.8 & 4.8$\pm$0.8 & 161.0$\pm$4.8 & 73.3$\pm$1.7 & 9.7$\pm$3.1 & \textless 11.4 & 34.3$\pm$4.1 & 24.0$\pm$3.8 & 10.2$\pm$0.3 & 5.7$\pm$0.1 \\
			0, 10 & 2.4$\pm$0.3 & 12.2$\pm$0.7 & 26.3$\pm$1.2 & 35.7$\pm$1.1 & 521.8$\pm$5.9 & 235.1$\pm$3.7 & 18.1$\pm$2.6 & 94.3$\pm$5.5 & 132.8$\pm$6.0 & 178.1$\pm$5.7 & 33.2$\pm$0.4 & 18.1$\pm$0.3 \\
			10, 10 & 1.8$\pm$0.4 & 5.5$\pm$0.6 & 20.5$\pm$0.8 & 26.8$\pm$1.0 & 544.2$\pm$6.2 & 264.4$\pm$3.5 & 13.5$\pm$2.8 & 42.7$\pm$4.3 & 103.9$\pm$4.1 & 133.6$\pm$5.2 & 34.6$\pm$0.4 & 20.4$\pm$0.3 \\
			20, 10 & 1.0$\pm$0.2 & 2.8$\pm$0.4 & 16.4$\pm$0.7 & 22.7$\pm$0.7 & 495.7$\pm$6.4 & 251.1$\pm$2.5 & 7.3$\pm$1.8 & 21.3$\pm$2.8 & 83.2$\pm$3.4 & 113.2$\pm$3.7 & 31.6$\pm$0.4 & 19.4$\pm$0.2 \\
			30, 10 & \textless 1.0 & \textless 2.3 & 7.6$\pm$0.7 & 12.0$\pm$0.9 & 256.9$\pm$5.7 & 154.9$\pm$2.5 & \textless 7.8 & \textless 18.0 & 38.7$\pm$3.8 & 60.0$\pm$4.3 & 16.4$\pm$0.4 & 11.9$\pm$0.2 \\
			40, 10 & \textless 1.1 & \textless 2.1 & 2.9$\pm$0.6 & 3.9$\pm$0.7 & 99.3$\pm$4.7 & 32.8$\pm$2.5 & \textless 8.4 & \textless 16.5 & 14.5$\pm$3.1 & 19.7$\pm$3.4 & 6.3$\pm$0.3 & 2.5$\pm$0.2 \\
			--10, 10 & 2.5$\pm$0.3 & 11.6$\pm$0.5 & 32.8$\pm$0.8 & 41.8$\pm$1.4 & 646.0$\pm$7.5 & 327.4$\pm$3.6 & 18.8$\pm$2.4 & 89.8$\pm$4.0 & 165.8$\pm$3.8 & 208.9$\pm$7.2 & 41.1$\pm$0.5 & 25.3$\pm$0.3 \\
			--20, 10 & 2.4$\pm$0.3 & 9.0$\pm$0.4 & 35.5$\pm$0.6 & 44.6$\pm$1.0 & 618.6$\pm$7.5 & 331.8$\pm$1.9 & 18.2$\pm$2.1 & 69.6$\pm$3.2 & 179.6$\pm$3.3 & 222.5$\pm$5.2 & 39.4$\pm$0.5 & 25.6$\pm$0.1 \\
			--30, 10 & \textless 1.5 & 4.6$\pm$0.8 & 19.7$\pm$0.6 & 24.4$\pm$0.9 & 425.2$\pm$4.7 & 203.6$\pm$4.0 & \textless 11.7 & 35.2$\pm$5.8 & 99.7$\pm$3.1 & 121.8$\pm$4.3 & 27.1$\pm$0.3 & 15.7$\pm$0.3 \\
			--40, 10 & \textless 1.1 & \textless 2.1 & 8.8$\pm$0.7 & 7.7$\pm$0.6 & 235.7$\pm$5.1 & 84.6$\pm$1.9 & \textless 8.1 & \textless 16.5 & 44.6$\pm$3.7 & 38.3$\pm$3.1 & 15.0$\pm$0.3 & 6.5$\pm$0.2 \\
			0, --10 & 3.5$\pm$0.4 & 9.6$\pm$0.6 & 37.0$\pm$1.0 & 55.6$\pm$1.3 & 472.9$\pm$7.2 & 485.2$\pm$3.2 & 26.4$\pm$3.1 & 73.7$\pm$5.0 & 186.9$\pm$5.1 & 277.5$\pm$6.5 & 30.1$\pm$0.5 & 37.4$\pm$0.2 \\
			10, --10 & 3.5$\pm$0.5 & 7.1$\pm$0.8 & 34.7$\pm$0.9 & 58.0$\pm$1.7 & 473.3$\pm$6.6 & 533.7$\pm$4.8 & 26.4$\pm$3.6 & 54.6$\pm$6.0 & 175.5$\pm$4.4 & 289.7$\pm$8.2 & 30.1$\pm$0.4 & 41.2$\pm$0.4 \\
			20, --10 & 1.9$\pm$0.3 & 4.5$\pm$0.3 & 22.5$\pm$0.6 & 37.6$\pm$1.1 & 367.4$\pm$7.4 & 406.1$\pm$2.4 & 14.4$\pm$2.2 & 35.0$\pm$2.4 & 113.9$\pm$3.3 & 187.7$\pm$5.7 & 23.4$\pm$0.5 & 31.3$\pm$0.2 \\
			30, --10 & \textless 1.6 & 5.9$\pm$0.5 & 8.0$\pm$0.8 & 14.9$\pm$1.0 & 232.8$\pm$7.4 & 214.3$\pm$2.9 & \textless 11.7 & 45.4$\pm$4.1 & 40.2$\pm$3.8 & 74.5$\pm$4.9 & 14.8$\pm$0.5 & 16.5$\pm$0.2 \\
			40, --10 & \textless 1.5 & \textless 1.8 & 6.5$\pm$1.1 & 6.1$\pm$1.2 & 140.9$\pm$6.1 & 77.5$\pm$3.4 & \textless 11.7 & \textless 14.1 & 32.7$\pm$5.3 & 30.5$\pm$6.1 & 9.0$\pm$0.4 & 6.0$\pm$0.3 \\
			--10, --10 & 2.3$\pm$0.3 & 4.0$\pm$0.5 & 26.1$\pm$1.0 & 43.0$\pm$1.3 & 420.2$\pm$7.8 & 467.3$\pm$3.7 & 17.2$\pm$1.9 & 30.9$\pm$3.9 & 132.2$\pm$5.2 & 214.9$\pm$6.6 & 26.8$\pm$0.5 & 36.0$\pm$0.3 \\
			--20, --10 & 2.3$\pm$0.4 & \textless 1.7 & 18.1$\pm$0.7 & 26.7$\pm$1.0 & 317.0$\pm$4.9 & 262.8$\pm$2.0 & 17.3$\pm$2.7 & \textless 13.5 & 91.4$\pm$3.8 & 133.4$\pm$4.7 & 20.2$\pm$0.3 & 20.3$\pm$0.2 \\
			--30, --10 & 1.1$\pm$0.2 & \textless 1.8 & 6.7$\pm$0.8 & 10.2$\pm$0.8 & 176.0$\pm$5.9 & 102.0$\pm$3.0 & 8.5$\pm$1.7 & \textless 14.1 & 33.7$\pm$4.2 & 51.0$\pm$4.1 & 11.2$\pm$0.4 & 7.9$\pm$0.2 \\
			--40, --10 & \textless 0.8 & \textless 1.9 & 3.6$\pm$0.8 & \textless 2.6 & 103.6$\pm$5.1 & 34.6$\pm$2.2 & \textless 6.3 & \textless 14.4 & 18.3$\pm$4.1 & \textless 13.2 & 6.6$\pm$0.3 & 2.7$\pm$0.2 \\
			0, 20 & 2.4$\pm$0.4 & 2.2$\pm$0.5 & 6.7$\pm$1.1 & 5.8$\pm$0.8 & 287.2$\pm$6.0 & 77.1$\pm$2.0 & 18.2$\pm$3.2 & 16.9$\pm$3.9 & 33.9$\pm$5.8 & 28.7$\pm$3.9 & 18.3$\pm$0.4 & 5.9$\pm$0.2 \\
			10, 20 & \textless 1.8 & 2.9$\pm$0.8 & 2.4$\pm$0.6 & 5.0$\pm$0.8 & 275.7$\pm$5.6 & 108.3$\pm$2.9 & \textless 14.1 & 22.3$\pm$6.5 & 12.0$\pm$3.1 & 25.2$\pm$3.8 & 17.6$\pm$0.4 & 8.3$\pm$0.2 \\
			20, 20 & \textless 1.1 & \textless 1.9 & 4.3$\pm$0.6 & 6.6$\pm$0.9 & 263.2$\pm$5.6 & 73.7$\pm$2.5 & \textless 8.1 & \textless 15.0 & 21.8$\pm$3.1 & 33.0$\pm$4.4 & 16.8$\pm$0.4 & 5.7$\pm$0.2 \\
			30, 20 & \textless 1.3 & \textless 2.2 & 2.8$\pm$0.7 & 3.4$\pm$0.6 & 162.7$\pm$5.5 & 61.9$\pm$2.0 & \textless 10.2 & \textless 17.4 & 14.0$\pm$3.3 & 17.0$\pm$3.1 & 10.4$\pm$0.4 & 4.8$\pm$0.2 \\
			40, 20 & \textless 1.2 & \textless 2.1 & \textless 2.1 & \textless 2.9 & 80.4$\pm$6.4 & 25.9$\pm$2.2 & \textless 9.3 & \textless 15.9 & \textless 10.5 & \textless 14.4 & 5.1$\pm$0.4 & 2.0$\pm$0.2 \\
			--10, 20 & 2.8$\pm$0.4 & 4.8$\pm$0.6 & 8.1$\pm$0.7 & 8.6$\pm$0.8 & 332.2$\pm$6.5 & 81.6$\pm$1.9 & 21.5$\pm$2.7 & 37.4$\pm$4.3 & 41.2$\pm$3.6 & 43.2$\pm$4.1 & 21.2$\pm$0.4 & 6.3$\pm$0.1 \\
			--20, 20 & 1.2$\pm$0.3 & 1.9$\pm$0.6 & 8.8$\pm$0.7 & 8.6$\pm$0.8 & 318.7$\pm$5.7 & 73.1$\pm$2.2 & 8.8$\pm$2.7 & 14.6$\pm$4.4 & 44.3$\pm$3.4 & 42.7$\pm$4.2 & 20.3$\pm$0.4 & 5.6$\pm$0.2 \\
			--30, 20 & \textless 1.2 & \textless 2.7 & 6.7$\pm$0.6 & 8.0$\pm$0.9 & 241.3$\pm$5.4 & 75.6$\pm$2.1 & \textless 9.3 & \textless 20.7 & 34.1$\pm$3.1 & 40.2$\pm$4.5 & 15.4$\pm$0.3 & 5.8$\pm$0.2 \\
			--40, 20 & 1.4$\pm$0.5 & \textless 2.3 & 3.0$\pm$0.4 & 3.3$\pm$0.6 & 180.1$\pm$4.7 & 47.9$\pm$2.2 & 11.0$\pm$3.6 & \textless 17.7 & 15.4$\pm$2.0 & 16.5$\pm$2.9 & 11.5$\pm$0.3 & 3.7$\pm$0.2 \\
			0, --20 & \textless 0.9 & 2.5$\pm$0.6 & 9.6$\pm$0.7 & 9.6$\pm$0.9 & 218.1$\pm$8.9 & 123.0$\pm$2.5 & \textless 7.2 & 19.7$\pm$4.7 & 48.4$\pm$3.5 & 48.1$\pm$4.6 & 13.9$\pm$0.6 & 9.5$\pm$0.2 \\
			10, --20 & \textless 1.8 & \textless 1.9 & 5.8$\pm$0.6 & 11.5$\pm$1.0 & 198.5$\pm$7.6 & 151.1$\pm$3.2 & \textless 13.5 & \textless 14.4 & 29.1$\pm$3.0 & 57.6$\pm$5.0 & 12.6$\pm$0.5 & 11.6$\pm$0.2 \\
			20, --20 & \textless 1.3 & \textless 2.1 & 2.5$\pm$0.5 & 11.6$\pm$1.2 & 185.6$\pm$8.9 & 101.1$\pm$3.6 & \textless 9.9 & \textless 15.9 & 12.6$\pm$2.4 & 58.0$\pm$6.1 & 11.8$\pm$0.6 & 7.8$\pm$0.3 \\
			30, --20 & \textless 1.3 & 2.9$\pm$0.9 & \textless 2.1 & 13.6$\pm$1.2 & 140.8$\pm$7.0 & 80.1$\pm$2.0 & \textless 9.6 & 22.5$\pm$6.8 & \textless 10.8 & 68.0$\pm$5.9 & 9.0$\pm$0.4 & 6.2$\pm$0.2 \\
			40, --20 & \textless 0.8 & 1.7$\pm$0.5 & 2.4$\pm$0.8 & 4.1$\pm$0.9 & 96.4$\pm$5.5 & 40.0$\pm$2.7 & \textless 6.3 & 12.9$\pm$3.9 & 11.9$\pm$4.0 & 20.3$\pm$4.5 & 6.1$\pm$0.3 & 3.1$\pm$0.2 \\
			--10, --20 & \textless 0.7 & \textless 1.4 & 5.3$\pm$0.7 & 9.8$\pm$0.9 & 171.5$\pm$6.0 & 125.6$\pm$2.5 & \textless 5.1 & \textless 11.1 & 26.5$\pm$3.4 & 48.7$\pm$4.4 & 10.9$\pm$0.4 & 9.7$\pm$0.2 \\
			--20, --20 & \textless 1.4 & \textless 2.2 & 2.3$\pm$0.6 & 5.2$\pm$0.6 & 128.8$\pm$4.3 & 79.5$\pm$1.8 & \textless 10.8 & \textless 16.8 & 11.5$\pm$2.8 & 25.9$\pm$3.2 & 8.2$\pm$0.3 & 6.1$\pm$0.1 \\
			--30, --20 & \textless 2.4 & \textless 2.1 & 5.1$\pm$0.7 & 4.7$\pm$0.8 & 83.0$\pm$4.9 & 45.4$\pm$2.2 & \textless 18.6 & \textless 16.5 & 25.8$\pm$3.3 & 23.4$\pm$4.0 & 5.3$\pm$0.3 & 3.5$\pm$0.2 \\
			--40, --20 & \textless 1.8 & \textless 2.3 & 4.4$\pm$1.1 & \textless 2.8 & 57.4$\pm$4.5 & 32.4$\pm$1.5 & \textless 13.5 & \textless 18.0 & 22.5$\pm$5.7 & \textless 14.1 & 3.7$\pm$0.3 & 2.5$\pm$0.1 \\
			\hline
			\vspace{0.01em} 
			\parbox{\textwidth}{\tiny \textbf{Note.} The uncertainties listed in this table represent measurement errors. For the positions without significant detections, we estimated a 3$\sigma$ upper limit to the line integrated intensities.}
		\end{tabular}
	\end{sidewaystable}

	\begin{sidewaystable}
		\scriptsize
		\caption{Integrated intensities and luminosities}
        \label{tab:A2}
		\begin{tabular}{*{13}{>{\centering\arraybackslash}p{1.5cm}}}
			\hline
			offset & 
			$I_{\rm HCN(4--3)}$ & 
			$I_{\rm HCO^+(4--3)}$ &
			$I_{\rm HCN(1--0)}$& 
			$I_{\rm HCO^+(1--0)}$ & 
			$I_{\rm CO(1--0)}$ &  
			$I_{\rm CO(3--2)}$ &
			$L'_{\rm HCN(4--3)}$ & 
			$L'_{\rm HCO^+(4--3)}$ &
			$L'_{\rm HCN(1--0)}$& 
			$L'_{\rm HCO^+(1--0)}$ & 
			$L'_{\rm CO(1--0)}$ & 
			$L'_{\rm CO(3--2)}$ \\
			(arcsec) &  
			(${\rm K\,km\,s^{-1}}$) &  
			(${\rm K\,km\,s^{-1}}$) &  
			(${\rm K\,km\,s^{-1}}$) &  
			(${\rm K\,km\,s^{-1}}$) &  
			(${\rm K\,km\,s^{-1}}$) &  
			(${\rm K\,km\,s^{-1}}$) &   
			(${\rm 10^4\,K\,km\,s^{-1}\,pc^{2}}$) & 
			(${\rm 10^4\,K\,km\,s^{-1}\,pc^{2}}$) & 
			(${\rm 10^4\,K\,km\,s^{-1}\,pc^{2}}$) & 
			(${\rm 10^4\,K\,km\,s^{-1}\,pc^{2}}$) & 
			(${\rm 10^6\,K\,km\,s^{-1}\,pc^{2}}$) & 
			(${\rm 10^6\,K\,km\,s^{-1}\,pc^{2}}$)\\    
			\hline
			0, 30 & \textless 1.0 & 3.1$\pm$0.6 & 2.3$\pm$0.7 & \textless 2.2 & 147.7$\pm$5.5 & 29.3$\pm$2.0 & \textless 8.1 & 24.3$\pm$4.4 & 11.8$\pm$3.7 & \textless 10.8 & 9.4$\pm$0.4 & 2.3$\pm$0.2 \\
			10, 30 & \textless 1.3 & \textless 2.2 & \textless 3.1 & \textless 3.1 & 154.3$\pm$6.1 & 42.9$\pm$2.3 & \textless 9.6 & \textless 16.8 & \textless 15.9 & \textless 15.6 & 9.8$\pm$0.4 & 3.3$\pm$0.2 \\
			20, 30 & \textless 1.2 & \textless 1.9 & \textless 1.9 & 3.8$\pm$0.9 & 130.1$\pm$5.6 & 19.9$\pm$1.3 & \textless 8.7 & \textless 14.4 & \textless 9.3 & 19.2$\pm$4.4 & 8.3$\pm$0.4 & 1.5$\pm$0.1 \\
			30, 30 & \textless 1.5 & \textless 1.8 & \textless 3.2 & 5.1$\pm$1.1 & 80.7$\pm$5.7 & 27.9$\pm$2.8 & \textless 11.4 & \textless 14.1 & \textless 15.9 & 25.3$\pm$5.7 & 5.1$\pm$0.4 & 2.2$\pm$0.2 \\
			40, 30 & \textless 1.2 & \textless 1.8 & \textless 2.1 & \textless 2.5 & 45.9$\pm$5.4 & 12.5$\pm$3.3 & \textless 9.0 & \textless 13.5 & \textless 10.5 & \textless 12.3 & 2.9$\pm$0.3 & 1.0$\pm$0.3 \\
			--10, 30 & \textless 1.2 & \textless 2.4 & 5.7$\pm$0.7 & \textless 2.5 & 167.3$\pm$8.4 & 39.6$\pm$2.6 & \textless 9.3 & \textless 18.9 & 29.0$\pm$3.4 & \textless 12.6 & 10.7$\pm$0.5 & 3.0$\pm$0.2 \\
			--20, 30 & \textless 1.2 & \textless 2.8 & 4.0$\pm$1.1 & \textless 2.7 & 149.2$\pm$5.2 & 16.4$\pm$1.9 & \textless 8.7 & \textless 21.6 & 20.2$\pm$5.7 & \textless 13.5 & 9.5$\pm$0.3 & 1.3$\pm$0.1 \\
			--30, 30 & \textless 1.4 & \textless 2.4 & 1.6$\pm$0.5 & \textless 2.8 & 130.8$\pm$5.2 & 25.4$\pm$2.4 & \textless 10.5 & \textless 18.6 & 8.3$\pm$2.8 & \textless 14.1 & 8.3$\pm$0.3 & 2.0$\pm$0.2 \\
			--40, 30 & \textless 1.1 & \textless 1.8 & \textless 1.9 & \textless 2.6 & 94.2$\pm$5.5 & 14.7$\pm$1.6 & \textless 8.1 & \textless 14.1 & \textless 9.6 & \textless 12.9 & 6.0$\pm$0.4 & 1.1$\pm$0.1 \\
			0, --30 & \textless 0.6 & \textless 1.8 & 3.3$\pm$0.8 & 6.3$\pm$0.6 & 113.6$\pm$5.2 & 55.9$\pm$1.8 & \textless 4.8 & \textless 14.1 & 16.9$\pm$4.0 & 31.5$\pm$3.1 & 7.2$\pm$0.3 & 4.3$\pm$0.1 \\
			10, --30 & \textless 1.0 & \textless 1.6 & 3.2$\pm$0.9 & 6.4$\pm$0.9 & 136.8$\pm$6.2 & 49.3$\pm$3.4 & \textless 7.8 & \textless 12.6 & 16.2$\pm$4.6 & 31.8$\pm$4.6 & 8.7$\pm$0.4 & 3.8$\pm$0.3 \\
			20, --30 & \textless 1.4 & \textless 2.6 & \textless 2.3 & \textless 3.4 & 129.9$\pm$6.9 & 53.6$\pm$3.0 & \textless 11.1 & \textless 20.4 & \textless 11.7 & \textless 17.1 & 8.3$\pm$0.4 & 4.1$\pm$0.2 \\
			30, --30 & \textless 1.4 & 3.8$\pm$0.8 & \textless 2.3 & 4.7$\pm$1.3 & 142.5$\pm$6.3 & 62.1$\pm$2.9 & \textless 11.1 & 29.0$\pm$5.8 & \textless 11.7 & 23.6$\pm$6.5 & 9.1$\pm$0.4 & 4.8$\pm$0.2 \\
			40, --30 & \textless 1.0 & 2.2$\pm$0.5 & 2.8$\pm$0.6 & \textless 1.9 & 109.9$\pm$5.3 & 45.7$\pm$1.9 & \textless 7.5 & 17.0$\pm$4.1 & 14.3$\pm$3.1 & \textless 9.3 & 7.0$\pm$0.3 & 3.5$\pm$0.1 \\
			--10, --30 & \textless 0.9 & 2.4$\pm$0.5 & 1.9$\pm$0.5 & 2.8$\pm$0.4 & 76.6$\pm$4.4 & 47.0$\pm$2.8 & \textless 7.2 & 18.2$\pm$3.8 & 9.5$\pm$2.6 & 14.1$\pm$2.1 & 4.9$\pm$0.3 & 3.6$\pm$0.2 \\
			--20, --30 & \textless 1.7 & 4.4$\pm$0.8 & \textless 2.9 & \textless 2.2 & 51.1$\pm$4.2 & 36.3$\pm$1.9 & \textless 13.2 & 34.0$\pm$6.1 & \textless 14.7 & \textless 11.1 & 3.3$\pm$0.3 & 2.8$\pm$0.1 \\
			--30, --30 & \textless 1.5 & 1.6$\pm$0.5 & \textless 1.5 & \textless 1.5 & 36.1$\pm$4.4 & 14.6$\pm$1.6 & \textless 11.4 & 12.4$\pm$3.9 & \textless 7.5 & \textless 7.5 & 2.3$\pm$0.3 & 1.1$\pm$0.1 \\
			--40, --30 & \textless 1.0 & 3.0$\pm$0.7 & \textless 1.7 & \textless 1.6 & 38.1$\pm$6.1 & 12.0$\pm$1.2 & \textless 7.5 & 23.2$\pm$5.1 & \textless 8.7 & \textless 7.8 & 2.4$\pm$0.4 & 0.9$\pm$0.1 \\
			0, 40 & \textless 1.3 & \textless 4.2 & \textless 2.8 & \textless 3.4 & 64.4$\pm$4.4 & 13.6$\pm$2.4 & \textless 9.9 & \textless 32.4 & \textless 14.1 & \textless 17.1 & 4.1$\pm$0.3 & 1.1$\pm$0.2 \\
			10, 40 & \textless 1.4 & \textless 2.8 & \textless 3.1 & \textless 2.3 & 70.0$\pm$4.8 & 17.7$\pm$2.6 & \textless 11.1 & \textless 21.6 & \textless 15.9 & \textless 11.1 & 4.5$\pm$0.3 & 1.4$\pm$0.2 \\
			20, 40 & \textless 1.0 & \textless 1.7 & \textless 2.1 & \textless 2.1 & 61.4$\pm$4.2 & 9.0$\pm$1.8 & \textless 7.5 & \textless 12.9 & \textless 10.8 & \textless 10.5 & 3.9$\pm$0.3 & 0.7$\pm$0.1 \\
			30, 40 & \textless 1.2 & \textless 2.0 & \textless 2.9 & \textless 2.9 & 43.3$\pm$4.3 & 14.8$\pm$2.3 & \textless 8.7 & \textless 15.3 & \textless 14.7 & \textless 14.7 & 2.8$\pm$0.3 & 1.1$\pm$0.2 \\
			40, 40 & \textless 1.1 & \textless 2.2 & --- & --- & 20.2$\pm$6.3 & \textless 7.2 & \textless 8.4 & \textless 17.1 & --- & --- & 1.3$\pm$0.4 & \textless 0.6 \\
			--10, 40 & \textless 1.2 & \textless 2.2 & \textless 2.7 & \textless 2.5 & 57.8$\pm$4.6 & 17.6$\pm$1.6 & \textless 9.3 & \textless 17.1 & \textless 13.5 & \textless 12.6 & 3.7$\pm$0.3 & 1.4$\pm$0.1 \\
			--20, 40 & \textless 0.9 & \textless 1.6 & \textless 3.0 & \textless 2.2 & 65.2$\pm$5.4 & \textless 3.8 & \textless 7.2 & \textless 12.3 & \textless 15.0 & \textless 11.1 & 4.1$\pm$0.3 & \textless 0.3 \\
			--30, 40 & \textless 1.5 & \textless 1.8 & \textless 2.4 & \textless 2.7 & 56.1$\pm$5.0 & 12.4$\pm$2.9 & \textless 11.4 & \textless 13.8 & \textless 12.0 & \textless 13.8 & 3.6$\pm$0.3 & 1.0$\pm$0.2 \\
			--40, 40 & \textless 0.8 & 1.8$\pm$0.6 & --- & --- & 50.3$\pm$4.3 & \textless 3.5 & \textless 6.0 & 13.7$\pm$4.3 & --- & --- & 3.2$\pm$0.3 & \textless 0.3 \\
			0, --40 & \textless 0.9 & 1.9$\pm$0.6 & 3.0$\pm$1.0 & 3.6$\pm$0.9 & 75.7$\pm$4.8 & 40.8$\pm$1.8 & \textless 7.2 & 14.9$\pm$4.9 & 15.2$\pm$5.0 & 17.9$\pm$4.2 & 4.8$\pm$0.3 & 3.2$\pm$0.1 \\
			10, --40 & \textless 1.3 & \textless 1.7 & \textless 3.3 & 5.1$\pm$0.8 & 67.5$\pm$5.0 & 39.2$\pm$1.9 & \textless 9.9 & \textless 12.9 & \textless 16.8 & 25.2$\pm$3.8 & 4.3$\pm$0.3 & 3.0$\pm$0.1 \\
			20, --40 & \textless 1.7 & \textless 2.2 & \textless 3.2 & \textless 3.4 & 86.6$\pm$7.0 & 31.5$\pm$1.7 & \textless 12.9 & \textless 16.8 & \textless 16.5 & \textless 17.1 & 5.5$\pm$0.4 & 2.4$\pm$0.1 \\
			30, --40 & \textless 1.2 & \textless 2.1 & \textless 2.0 & \textless 2.6 & 89.5$\pm$4.5 & 31.1$\pm$1.8 & \textless 9.0 & \textless 15.9 & \textless 9.9 & \textless 12.9 & 5.7$\pm$0.3 & 2.4$\pm$0.1 \\
			40, --40 & \textless 1.2 & \textless 1.5 & --- & --- & 93.6$\pm$5.0 & 30.3$\pm$1.7 & \textless 9.0 & \textless 11.7 & --- & --- & 6.0$\pm$0.3 & 2.3$\pm$0.1 \\
			--10, --40 & \textless 0.7 & \textless 1.6 & \textless 2.0 & \textless 2.0 & 58.9$\pm$4.1 & 16.8$\pm$1.9 & \textless 5.7 & \textless 12.6 & \textless 10.2 & \textless 9.9 & 3.7$\pm$0.3 & 1.3$\pm$0.1 \\
			--20, --40 & \textless 1.6 & \textless 2.4 & \textless 2.1 & \textless 2.0 & 32.7$\pm$3.6 & 13.4$\pm$1.5 & \textless 12.3 & \textless 18.3 & \textless 10.8 & \textless 10.2 & 2.1$\pm$0.2 & 1.0$\pm$0.1 \\
			--30, --40 & 1.8$\pm$0.4 & \textless 1.8 & \textless 2.1 & \textless 2.1 & 17.8$\pm$4.0 & 10.4$\pm$1.7 & 13.6$\pm$3.2 & \textless 14.1 & \textless 10.5 & \textless 10.5 & 1.1$\pm$0.3 & 0.8$\pm$0.1 \\
			--40, --40 & \textless 1.0 & 1.9$\pm$0.4 & --- & --- & 17.5$\pm$3.3 & 5.4$\pm$1.1 & \textless 7.2 & 14.3$\pm$2.8 & --- & --- & 1.1$\pm$0.2 & 0.4$\pm$0.1 \\
			\hline
			\vspace{0.01em} 
			\parbox{\textwidth}{\tiny \textbf{Note.} The uncertainties listed in this table represent measurement errors. For the positions without significant detections, we estimated a 3$\sigma$ upper limit to the line integrated intensities.}
		\end{tabular}
	\end{sidewaystable}

		\begin{table}[h] 
			\centering
			\caption{Luminosity ratios at positions with detected \Jt{4}{3} emission}
			\label{tab:Luminosity Ratios}
			\begin{tabular}{ccccc}
				\hline
				offset  &  HCN(4--3)/HCN(1--0) & HCO$^+$(4--3)/HCO$^+$(1--0) &  HCN(4--3)/HCO$^+$(4--3) &  HCN(1--0)/HCO$^+$(1--0)    \\
				\hline
				0, 0 & 0.22 $\pm$ 0.01 & 0.43 $\pm$ 0.02 & 0.35 $\pm$ 0.02 & 0.67 $\pm$ 0.01 \\
				10, 0 & 0.17 $\pm$ 0.01 & 0.43 $\pm$ 0.02 & 0.27 $\pm$ 0.02 & 0.66 $\pm$ 0.02 \\
				$-$10, 0 & 0.14 $\pm$ 0.01 & 0.36 $\pm$ 0.01 & 0.26 $\pm$ 0.02 & 0.68 $\pm$ 0.02 \\
				20, 0 & 0.10 $\pm$ 0.01 & 0.30 $\pm$ 0.02 & 0.23 $\pm$ 0.02 & 0.69 $\pm$ 0.02 \\
				$-$20, 0 & 0.12 $\pm$ 0.01 & 0.17 $\pm$ 0.01 & 0.49 $\pm$ 0.05 & 0.72 $\pm$ 0.02 \\
				0, 10 & 0.14 $\pm$ 0.02 & 0.53 $\pm$ 0.04 & 0.19 $\pm$ 0.03 & 0.75 $\pm$ 0.04 \\
				10, 10 & 0.13 $\pm$ 0.03 & 0.32 $\pm$ 0.04 & 0.32 $\pm$ 0.07 & 0.78 $\pm$ 0.04 \\
				$-$10, 10 & 0.11 $\pm$ 0.02 & 0.43 $\pm$ 0.02 & 0.21 $\pm$ 0.03 & 0.79 $\pm$ 0.03 \\
				20, 10 & 0.09 $\pm$ 0.02 & 0.19 $\pm$ 0.03 & 0.34 $\pm$ 0.10 & 0.74 $\pm$ 0.04 \\
				$-$20, 10 & 0.10 $\pm$ 0.01 & 0.31 $\pm$ 0.02 & 0.26 $\pm$ 0.03 & 0.81 $\pm$ 0.02 \\
				0, $-$10 & 0.14 $\pm$ 0.02 & 0.27 $\pm$ 0.02 & 0.36 $\pm$ 0.05 & 0.67 $\pm$ 0.02 \\
				10, $-$10 & 0.15 $\pm$ 0.02 & 0.19 $\pm$ 0.02 & 0.48 $\pm$ 0.08 & 0.61 $\pm$ 0.02 \\
				$-$10, $-$10 & 0.13 $\pm$ 0.02 & 0.14 $\pm$ 0.02 & 0.56 $\pm$ 0.09 & 0.62 $\pm$ 0.03 \\
				20, $-$10 & 0.13 $\pm$ 0.02 & 0.19 $\pm$ 0.01 & 0.41 $\pm$ 0.07 & 0.61 $\pm$ 0.03 \\
				0, 20 & 0.54 $\pm$ 0.13 & 0.59 $\pm$ 0.16 & 1.07 $\pm$ 0.31 & 1.18 $\pm$ 0.26 \\
				$-$10, 20 & 0.52 $\pm$ 0.08 & 0.87 $\pm$ 0.13 & 0.57 $\pm$ 0.10 & 0.95 $\pm$ 0.12 \\
				$-$20, 20 & 0.20 $\pm$ 0.06 & 0.34 $\pm$ 0.11 & 0.60 $\pm$ 0.26 & 1.04 $\pm$ 0.13 \\
				\hline
				Mean & 0.18 $\pm$ 0.04 & 0.32 $\pm$ 0.06 & 0.39 $\pm$ 0.11 & 0.74 $\pm$ 0.08 \\
				\hline
			\end{tabular}
		\end{table}

		\begin{table}[h] 
			\centering
			\caption{Velocity-integrated intensities of emission lines at the SW and NE lobes, along with their corresponding SW-to-NE intensity ratios}
			\label{tab:intensity}
			\begin{tabular}{l c c c c c l}
				\hline
				\multirow{2}{*}{Line} & \multicolumn{2}{c}{Intensity [K km s$^{-1}$]} & \multirow{2}{*}{Beam [\arcsec]} & \multirow{2}{*}{SW/NE} & \multirow{2}{*}{Ref.} \\
				\cline{2-3}
				& NE lobe  & SW lobe & & & \\
				\hline
				CO(1--0)  	     & 755.3  $\pm$ 8.1  & 757.6 $\pm$ 7.5  & 14 & 1.00 $\pm$ 0.02 & This work \\
				& 573.9 $\pm$ 0.6  & 672.8 $\pm$ 0.8 & 22   & 1.17 $\pm$ 0.01 & \cite{Mao_2000} \\
				CO(2-1)          & 674.3 $\pm$ 0.8  & 804.9 $\pm$ 1.1 & 22   & 1.19 $\pm$ 0.01 & \cite{Mao_2000} \\
				CO(3--2)         & 656.5 $\pm$ 6.1 & 736.9 $\pm$ 3.2 & 14   & 1.12 $\pm$ 0.02 & This work\\
				& 523 $\pm$ 41.8    & 476 $\pm$ 38.1   & 24.4 & 0.91 $\pm$ 0.11 & \cite{Ward_2003} \\
				CO(4--3)         & 465.3 $\pm$ 12.9  & 503.0 $\pm$ 12.6 & 22   & 1.08 $\pm$ 0.04 & \cite{Mao_2000} \\
				& 427 $\pm$ 42.7  & 332 $\pm$ 33.2   & 24.4 & 0.78 $\pm$ 0.11 & \cite{Ward_2003} \\
				CO(5--4)         & 104.9 $\pm$ 1.2    & 135.1 $\pm$ 1.2 & 38   & 1.29 $\pm$ 0.02 & \cite{loenen_2010} \\
				CO(6--5)         & 99.4 $\pm$ 2.4   & 135.1 $\pm$ 2.4   & 33   & 1.36 $\pm$ 0.04 & \cite{loenen_2010} \\
				& 179 $\pm$ 21.5    & 224 $\pm$ 26.9   & 24.4 & 1.25 $\pm$ 0.21 & \cite{Ward_2003} \\
				CO(7--6)         & 132.3 $\pm$ 5.8  & 167.2 $\pm$ 6.4 & 22   & 1.26 $\pm$ 0.08 & \cite{Mao_2000} \\
				& 93.2 $\pm$ 4.9 & 124.1 $\pm$ 5.2   & 27   & 1.33 $\pm$ 0.09 & \cite{loenen_2010} \\
				CO(8--7)         & 80.7 $\pm$ 8.3 & 103.3 $\pm$ 7.0   & 25   & 1.28 $\pm$ 0.19 & \cite{loenen_2010} \\
				CO(9--8)         & 55.1 $\pm$ 4.3  & 96.0 $\pm$ 4.3 & 23   & 1.74 $\pm$ 0.18 & \cite{loenen_2010} \\
				CO(10--9)        & 59.6 $\pm$ 13.2 & 51.7 $\pm$ 11.3 & 20   & 0.87 $\pm$ 0.27 & \cite{loenen_2010} \\
				$^{13}$CO(1--0)  & 40.1 $\pm$ 0.5   & 58.8 $\pm$ 0.6  & 22   & 1.47 $\pm$ 0.03 & \cite{Mao_2000} \\
				$^{13}$CO(2-1)  & 58.9 $\pm$ 1.1   & 69.0 $\pm$ 1.1  & 22   & 1.17 $\pm$ 0.03 & \cite{Mao_2000} \\
				$^{13}$CO(3--2)  & 51.2 $\pm$ 4.9   & 68.0 $\pm$ 2.8  & 22   & 1.33 $\pm$ 0.14 & \cite{Mao_2000} \\
				$^{13}$CO(5--4)  & 4.0 $\pm$ 0.3    & 6.1 $\pm$ 0.3  & 44   & 1.53 $\pm$ 0.18 & \cite{loenen_2010} \\
				$^{13}$CO(6--5)  & 2.4 $\pm$ 0.6    & 7.0 $\pm$ 0.9   & 33   & 2.82 $\pm$ 0.88 & \cite{loenen_2010} \\
				HCN(1--0)  	     & 52.8 $\pm$ 1.0   & 62.9 $\pm$ 1.0   & 14 & 1.19 $\pm$ 0.03 & This work \\
				HCO$^+$(1--0) & 81.2 $\pm$ 1.9    & 93.9 $\pm$  1.6   & 14 & 1.16 $\pm$ 0.03 & This work \\
				HCN(3--2)  	     & 14.6 $\pm$ 3.0   & 15.1 $\pm$ 2.3  & 14 & 1.03 $\pm$ 0.28 & \cite{Wild_1992} \\
				HCO$^+$(3--2) & 23.2 $\pm$ 3.0  & 35 $\pm$ 4.0  & 14 & 1.51 $\pm$ 0.27 & \cite{Wild_1992} \\
				HCN(4--3)  	     & 6.1 $\pm$ 0.3    & 5.7 $\pm$ 0.3   & 14 & 0.93 $\pm$ 0.07 & This work \\
				HCO$^+$(4--3) & 22.6 $\pm$ 0.6 & 21.6 $\pm$ 0.6   & 14 & 0.96 $\pm$ 0.04 & This work \\
				\hline
			\end{tabular}
			\vspace{2mm}
			\begin{minipage}{0.85\textwidth}
				\footnotesize \textbf{Note.} The uncertainties from this work, \cite{Mao_2000}, \cite{loenen_2010}, and \cite{Wild_1992} represent measurement errors. The uncertainties from \cite{Ward_2003} are estimates based on a combination of factors, including effects related to calibration and smoothing.
			\end{minipage}
		\end{table}

\end{appendix}
	
\end{document}